\documentclass[fleqn,usenatbib]{mnras}

\usepackage{newtxtext,newtxmath}
\usepackage[T1]{fontenc}
\usepackage{ae,aecompl}

\usepackage{graphicx}	% Including figure files
\usepackage{amsmath}	% Advanced maths commands
\usepackage{amssymb}	% Extra maths symbols

\def\d{$^\circ$}

\newcommand{\kms}{\mathrm{km}\,\mathrm{s}^{-1}}
\newcommand{\msun}{\mathrm{M}_\odot}
\newcommand{\rsun}{\mathrm{R}_\odot}
\usepackage[usenames,dvipsnames]{xcolor}

\newcommand{\targ}{V1460~Her}

\title[\targ: A fast spinning white dwarf accreting from an evolved donor star]{\targ: A fast spinning white dwarf accreting from an evolved donor star}

\author[R.P. Ashley et al.]{R.\ P.\ Ashley,$^{1,2}$\thanks{E-mail: r.p.ashley@warwick.ac.uk}
T.\ R.\ Marsh,$^{1}$ 
E.\ Breedt,$^{3}$
B.\ T.\ G\"{a}nsicke,$^{1}$ 
A.\ F.\ Pala, $^{4}$ 
O.\ Toloza, $^{1}$
\newauthor P.\ Chote,$^{1}$
John\ R.\ Thorstensen,$^{5}$
M.\ R.\ Burleigh\,$^{6}$
\\
$^{1}$Department of Physics, University of Warwick, Gibbet Hill Road, Coventry, CV4 7AL, UK\\
$^{2}$Isaac Newton Group of Telescopes, Apartado de Correos 321, Santa Cruz de La Palma, E-38700, Spain\\
$^{3}$Institute of Astronomy, University of Cambridge, Madingley Road, Cambridge CB3 0HA, UK\\
$^{4}$European Southern Observatory, Karl Schwarzschild Stra{\ss}e 2, Garching, 85748, Germany\\
$^{5}$Dartmouth College Department of Physics and Astronomy, 6127 Wilder Laboratory, Hanover, NH 03755-3528,  USA\\
$^{6}$ Department of Physics and Astronomy, University of Leicester, Leicester, LE1 7RH, UK\\
}

\date{Accepted XXX. Received YYY; in original form ZZZ}

\pubyear{2020}

\hypersetup{draft}
\begin{document}
\label{firstpage}
\pagerange{\pageref{firstpage}--\pageref{lastpage}}
\maketitle

\begin{abstract}
We present time-resolved optical and ultraviolet spectroscopy and photometry of V1460~Her, an eclipsing cataclysmic variable with a 4.99\,h orbital period and an overluminous K5-type donor star. The optical spectra show emission lines from an accretion disc along with absorption lines from the donor. We use these to measure radial velocities, which, together with constraints upon the orbital inclination from photometry, imply masses of $M_1=0.869\pm0.006\,\msun$ and $M_2=0.295\pm0.004\,\msun$ for the white dwarf and the donor. The radius of the donor, $R_2=0.43\pm0.002\,\rsun$, is $\approx 50$ per cent larger than expected given its mass, while its spectral type is much earlier than the M3.5 type that would be expected from a main sequence star with a similar mass. \textit{HST} spectra show strong \ion{N}{v} 1240\,\AA\ emission but no \ion{C}{iv} 1550\,\AA\ emission, evidence for CNO-processed material. The donor is therefore a bloated, over-luminous remnant of a thermal-timescale stage of high mass transfer and has yet to re-establish thermal equilibrium. Remarkably, the \textit{HST} ultraviolet data also show a strong 30 per cent peak-to-peak, $38.9\,$s pulsation that we explain as being due to the spin of the white dwarf, potentially putting V1460~Her in a similar category to the propeller system AE~Aqr in terms of its spin frequency and evolutionary path. AE~Aqr also features a post-thermal timescale mass donor, and V1460~Her may therefore be its weak magnetic field analogue since the accretion disc is still present, with the white dwarf spin-up a result of a recent high accretion rate.
\end{abstract}


%%%%%%%%%%%%%%%%%%%%%%%%%%%%%%%%%%%%%%%%%%%%%%%%%%

%%%%%%%%%%%%%%%%% BODY OF PAPER %%%%%%%%%%%%%%%%%%

\section{Introduction}

Cataclysmic variable stars (CVs) contain white dwarf primary stars usually accreting from hydrogen-rich main-sequence-like secondaries. In most cases the accretion occurs via a disc and they are the closest and most easily observed examples of accretion onto compact objects. As CVs evolve, their secondary stars continually lose mass and therefore need to adjust their structures to maintain thermal equilibrium. Whether they manage to do so depends upon the thermal timescale of the secondary star compared to the mass loss timescale, $M_2/-\dot{M_2}$, where $M_2$ is the mass of the secondary star. Both timescales tend to lengthen as orbital periods and secondary star masses decline, but at very low masses ($< 0.1\,\msun$) and correspondingly short periods ($\sim 80\,$min), the battle to maintain thermal equilibrium is well and truly lost resulting in a minimum orbital period for systems with hydrogen-rich donors, \citep{Paczynski1981, GaensickeMinimum}. Until this point however, we expect the secondary stars to be fairly close to thermal equilibrium (although departure from thermal equilibrium is thought to be partly responsible for the paucity of systems between 2 and 3 hours known as the period gap). The maintenance of near thermal equilibrium while losing mass, and the one-to-one relationship between orbital period and the mean density of Roche-lobe filling stars \citep{1972ApJ...175L..79F} results in a strong correlation between orbital period, mass and spectral type amongst CV donor stars, \citep{2006MNRAS.373..484K}.  

The discovery of K-type donor stars in the short-period CVs EI\,Psc, \citep{2002ApJ...567L..49T} and QZ\,Ser \citep{2002PASP..114.1117T}, with periods of 1.1 and 2.0\,h, and K4 and K5 donors respectively, forcefully demonstrated that alternative evolution channels also contribute to the population of CVs. \citet{2002ApJ...567L..49T,2002PASP..114.1117T} argued that the over-luminous donors in EI\,PSc and QZ\,Ser were the stripped cores of companions that were initially more massive, $\ga1\,\msun$, bearing the signatures of CNO burning in the form of enhanced sodium abundances. Anomalous carbon and nitrogen abundances were confirmed in EI\,PSc in the far-ultraviolet \citep{Gaensicke2003} and infrared \citep{2016ApJ...833...14H}. In the ultraviolet, these anomalies are reflected in an inversion of the flux ratio of the \ion{N}{v} and \ion{C}{iv} resonance lines. In CVs with (quasi) main-sequence donors, $\ion{N}{v}/\ion{C}{iv}\simeq0.5$, whereas the \ion{C}{iv} line is undetected in the \textit{Hubble Space Telescope} (\textit{HST}) ultraviolet spectrum of EI\,Psc, with a lower limit of $\ion{N}{v}/\ion{C}{iv}\ga8$ \citep{Gaensicke2003}. \citet{2016ApJ...833...14H} report that the infrared spectra of both QZ\,Ser and EI\,Psc suggest an H deficiency of about 30\,per\,cent, and a depletion of C based on the weak or absent CO absorption bands~--~corroborating the abundance anomalies detected in the ultraviolet spectroscopy of EI\,Psc. Sufficiently high initial donor star masses, i.e. $M_2/M_1\ga1$, where $M_1$ is the mass of the primary, lead to thermal-timescale mass transfer, with a mass transfer rate that may result in stable shell burning on the white dwarf \citep{1982ApJ...257..767F, 1982ApJ...259..244I, 2013ApJ...777..136W}. During this shell-burning phase the systems will be luminous supersoft X-ray sources \citep{1992A&A...262...97V} and the white dwarf may grow in mass, making this evolutionary channel a potential single-degenerate pathway to type~Ia supernovae (SN\,Ia) \citep{2010ApJ...712..728D}. In many cases however, the rapid mass loss of the donor star results in such a decrease of the mass ratio that, failing to reach the ignition conditions for a thermonuclear supernova, the systems will blend into the population of normal CVs once the mass ratio is $M_2/M_1\la1$. 

Whereas the physical properties of both stellar components of these ``failed SN\,Ia'' can be accurately measured, providing a powerful diagnostic of this evolutionary pathway, only a relatively small number of such systems have so far been identified \citep[e.g.][]{2009A&A...496..805R, 2015PASP..127..351T, 2015AJ....149..128T, 2019MNRAS.489.1023Y, 2014ApJ...790...28R}. Here we report that the system \targ\ is the newest member of this exclusive list of CVs.

%It was thus a surprise when systems were discovered with donors that were much hotter than expected given their orbital period. There is now a small, but growing list of these \citep{2015PASP..127..351T, 2015AJ....149..128T}. 

%\textit{Steven's pre-supersoft X-ray source, some reference to SNIa models.} 

On 2016 June 4, \citet{2016ATel.9112....1T} reported that the Catalina Real-Time Transient Survey (CRTS,  \citealt{CatalinaCatalog}) had detected an outburst of CRTS J162117.3+441254, an object that had previously been classified as a contact binary in the SuperWASP \citep{2006PASP..118.1407P} survey with an orbital period of $0.207852\,$d 
\citep{2013A&A...549A..86L}.  \citet{2016ATel.9141....1T} obtained a full orbit of time-resolved optical spectroscopy in quiescence, which showed strong Balmer emission lines, and measured the radial velocities of absorption features in anti-phase with the emission, confirming the binary nature of the target and evidence that the donor was a K-type. It was clear that this was a CV undergoing a dwarf nova outburst.
\citet{2017NewA...52....8K} took photometric data two weeks after the outburst and supplemented this with AAVSO observations to make a qualitative model of the system. This model resulted in an orbital inclination of $i = 88 \pm3$\d, a secondary temperature of $T_2 = 4400 \pm 150$K and a relative radius with respect to the semi-major axis of $r_2 = 0.237 a$. They also showed that their model, during outburst, fits a disc of temperature $T_1 = 7600 \pm 200$\,K and, during quiescence, a spherical body of temperature $T_1 = 4400 \pm 150$\,K.  The outburst disc was modelled to have a radius of $r_1 = 0.36 \pm 0.01 a$. \citet{2019IBVS.6261....1K} designated this target with the General Catalogue of Variable Stars name of \targ\ and classified it as an eclipsing dwarf nova. 

In this paper we present the results of extensive optical and ultraviolet spectroscopic and photometric follow-up of \targ\ which we use to determine system parameters and the nature of its donor star. Serendipitously, our data also reveal rapid spin of its white dwarf. We begin with a description of our observations.

\section{Observations}
\label{observations}
Following the Astronomer's Telegram 9141 \citep{2016ATel.9141....1T} on 2016 June 11, we decided to take follow-up observations using the Warwick 1~metre (W1m) telescope on the island of La Palma. We maintained observations for several nights as the target evolved from outburst to quiescence. In addition, we submitted proposals and were granted time for spectroscopic follow-up with the William Herschel Telescope (WHT) on La Palma and the \textit{HST}. The details of these campaigns are summarised in the following sections. To supplement our own observations we made use of various other sources of photometry, including AAVSO, SuperWASP, ASAS-SN \citep{ASASSN1} and CRTS, see section \ref{sec:archive}. 

\subsection{Warwick One metre (W1m) photometry}
We used the W1m telescope to perform follow-up observations of the 2016 June outburst. The W1m is a robotic F/7 equatorial fork-mounted telescope with a dual-beam camera system. It has a dual-channel red-blue photometer with a dichroic filter to split the beam into separate detectors and is capable of taking simultaneous exposures in each band. The blue channel contains a BG40 filter which covers wavelengths from 3000 to 6000\,\AA, and the red channel a $Z$-band filter covering 8200 to 9300\,\AA. For four nights from the 2016 June 5 to 9, photometry of \targ\ was taken with the intention of covering as much of a full orbit as possible. A log of observations is provided in Table~\ref{tab:observations}. The W1m was undergoing a commissioning phase at the time and we used \targ\ as a serendipitous test target for the development of the observation pipeline. 

For all reductions, the same comparison star was used, this was TYC~3068-831-1 which is a magnitude V=12 star located about nine~arcmin to the south west of the target. Data was reduced using the  {\sc tsreduce} software \citep{tsreduce}. For the photometric extraction we used synthetic apertures with variable radii adapted to fit the seeing conditions on each frame. Since our blue filter is non-standard, we did not calibrate our data to an apparent magnitude scale and our analysis uses the relative flux of target to comparison.

\begin{table*}
  \caption{Log of our observations of \targ\ taken with the W1m, WHT and \textit{HST}. The W1m data were obtained immediately after the report of the 2016 June outburst of \targ\ with one additional night later  during quiescence, 44 days later. The number N denotes the approximate number of orbital cycles since the reported outburst. }
  \begin{tabular}{ l l l l l l l l}
  \hline
  Date & Instrument & Filters or & JD (start)                   & State & N        & $T_{\mbox{exp}}$\    & Duration \\
       &            & Grating (central wavelength \AA)        & &          & (orbits) &  (s)                 &  (hr)    \\
  \hline
    2016/06/05 & W1m & BG40 + Z & 2457545.386138 & Outburst & 21  & 30 & 5.68 \\
    2016/06/07 & W1m & BG40 + Z & 2457547.385385 & Outburst & 30  & 30 & 4.33 \\
    2016/06/08 & W1m & BG40 + Z & 2457548.528484 & Outburst & 36  & 30 & 4.63 \\
    2016/06/09 & W1m & BG40 + Z & 2457549.529722 & Outburst & 40  & 30 & 4.52 \\
    2016/07/23 & W1m & BG40 + Z & 2457593.373344 & Quiescence & 251 & 30 & 5.39 \\
    2016/07/11 & WHT ISIS & R1200B(4601) + R1200R(6400) & 2457581.574537 & Quiescence & 194 & 150 & 2.25  \\
    2016/08/16 & WHT ISIS & R1200B(4601) + R1200R(6400) & 2457617.363687 & Quiescence & 366 & 150 & 3.50 \\
    2017/06/20 & WHT ISIS & R1200B(4601) + R1200R(6400) & 2457925.372561 & Quiescence & 1848 & 150 & 6.37 \\
    2017/06/21 & WHT ISIS & R1200B(4601) + R1200R(6400) & 2457926.374255 & Quiescence & 1853 & 150 & 5.95 \\
    2017/07/03 & WHT ISIS & R1200B(4601) + R1200R(6400) & 2457938.414351 & Quiescence & 1911 & 150 & 4.45 \\
    2017/07/04 & WHT ISIS & R1200B(4601) + R1200R(6400) & 2457939.407028 & Quiescence & 1916 & 150 & 3.20  \\
    2017/03/02 & \textit{HST} COS & G140L (1105) & 2457815.429107 & Quiescence & 1319 & -- & 2.30 \\
  \hline
  \end{tabular}
  \label{tab:observations}
\end{table*}

\subsection{Published survey photometry}
\label{sec:archive}
\begin{figure*}
\centering
\includegraphics[width=\textwidth]{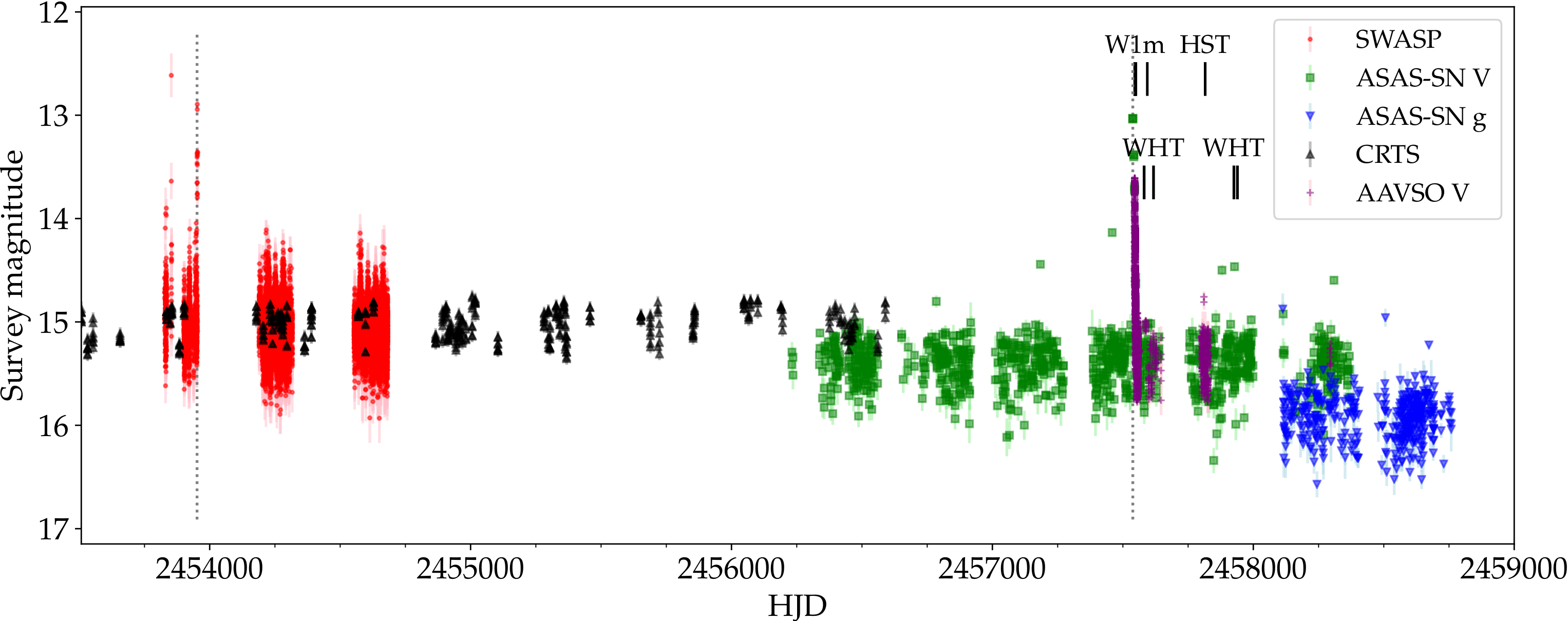}
\caption{SuperWASP, CRTS, ASAS-SN and AAVSO photometry for V1460 Her from 2005 December 5 to 2019 October 2. Evidence for outbursts appear at JD~2453952 and JD~2457538 (as indicated by the vertical dotted lines). The second outburst was also covered extensively by AAVSO observers and we show their V-band photometry here.  The surveys have different bandpasses and therefore their magnitude zeropoints differ. The total timespan is 5250\,d. Vertical marks indicate the dates of the W1m photometry (see Fig.~\ref{fig:stacked}, \ref{fig:eclipse}), the \textit{HST} ultraviolet spectroscopy (see Fig.~\ref{fig:hstspectrum}, \ref{fig:hst_lightcurve}, \ref{fig:hst_zoom}) and the WHT ISIS spectroscopy (see Fig.~\ref{fig:combined_spectra}, \ref{fig:templates}, \ref{fig:redtrails}, \ref{fig:halpha_subtracted}).  }
\label{fig:photometry_all}
\end{figure*}

\begin{figure}
\centering
\includegraphics[width=\columnwidth]{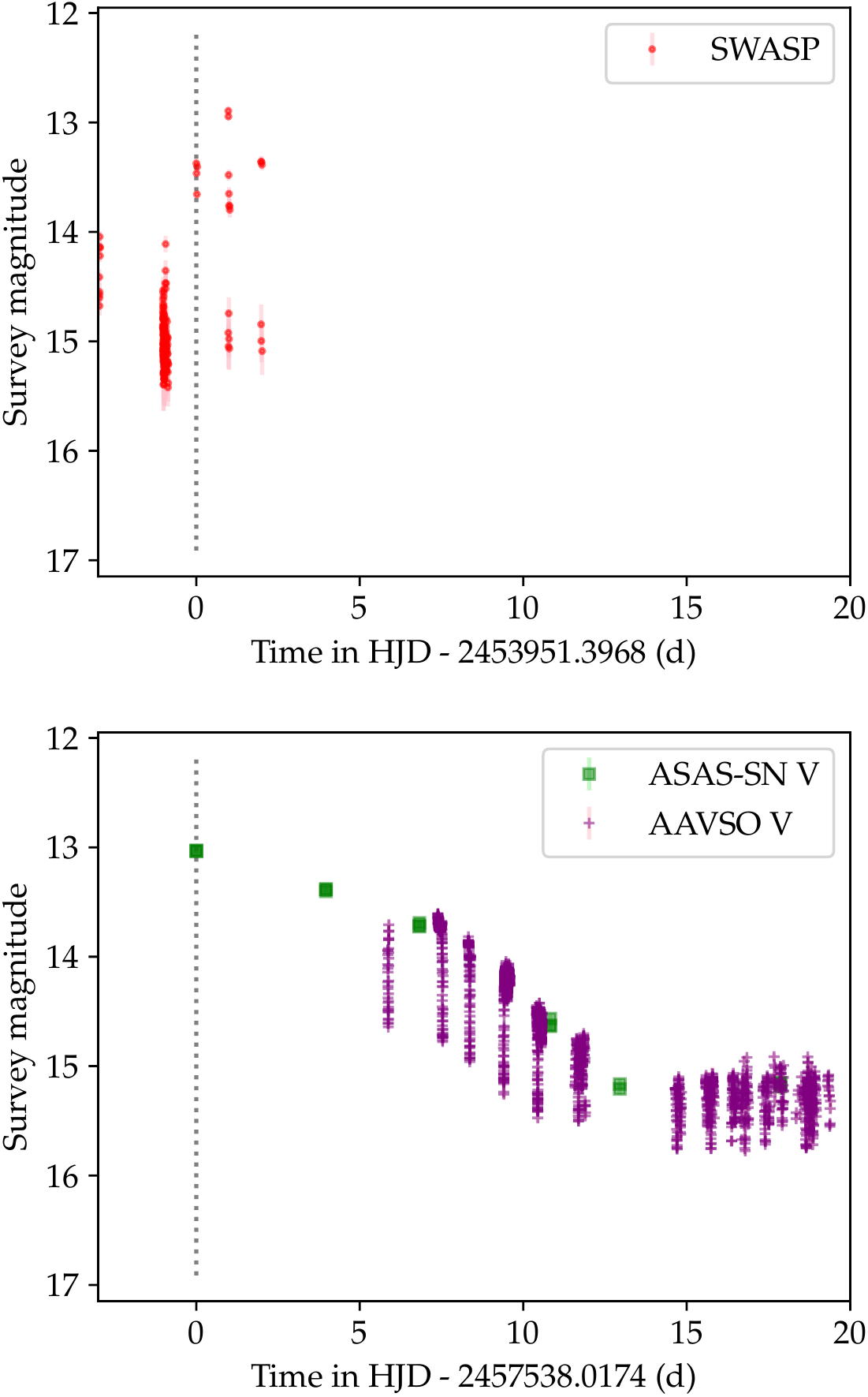}
\caption{SuperWASP, AAVSO and ASAS-SN photometry showing the outbursts of \targ. We plot all survey data from 3 days prior to the first detection of the outburst and continue for 20 days post detection. The AAVSO data covers eclipses during the orbit, showing how the eclipse depth progresses after outburst. }
\label{fig:outbursts}
\end{figure}

\targ\ appears to have displayed at least two outbursts in the last fourteen years, the discovery outburst that occurred in 2016 June (JD~2457538) and one recorded in the SuperWASP data in 2006 August (JD~2453952). These two outbursts are indicated with dotted lines in Fig. \ref{fig:photometry_all} and a closer examination of the outburst and return to quiescence is shown in Fig. \ref{fig:outbursts}. The SuperWASP data is presented in \citet{2013A&A...549A..86L} where they quote an orbital period of 0.207852(1)\,d, and classify the target as a contact binary of type W~UMa. 

\begin{figure}
\centering
\includegraphics[width=\columnwidth]{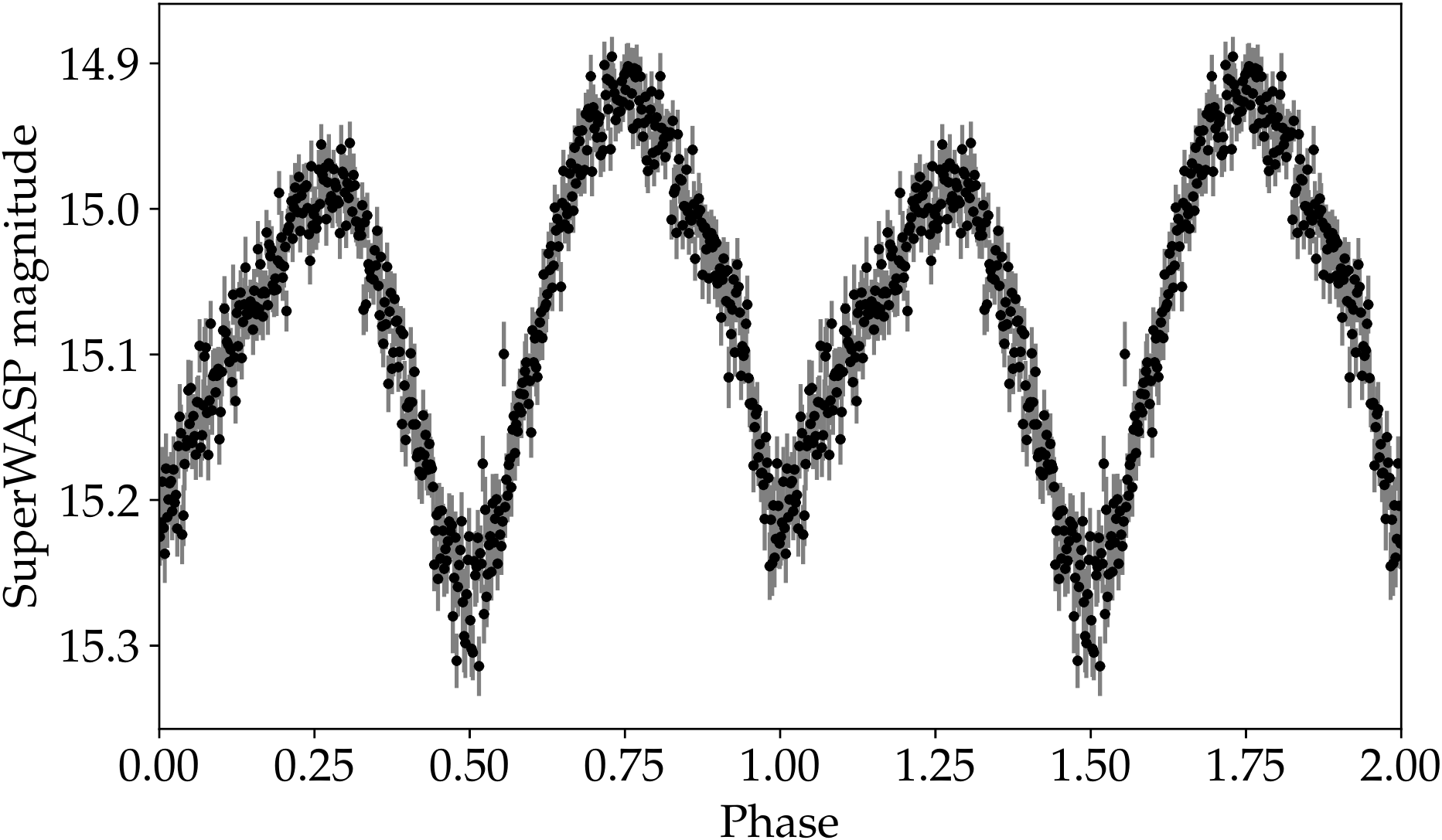}
\caption{Phase folded SuperWASP photometry for \targ. To create this figure, all of the SuperWASP data shown in Fig.~\ref{fig:photometry_all} have been phase-folded and then binned (with weighted average) into 500 phase bins. The ephemeris used to phase fold the data is given in equation \ref{eq:ephemeris}.}
\label{fig:bin_swasp}
\end{figure}

This is a system where the secondary star dominates the flux in the visible and near infrared part of the spectrum. The phase-folded light curve (Fig.~\ref{fig:bin_swasp}) of the SuperWASP data therefore shows modulation caused by the tidal distortion of the secondary star and gravity darkening on its near side. This light curve was used as a modulation to apply to the flux contribution of the secondary star in the time resolved spectroscopy study in section~\ref{spectroscopy}. The non-sinusoidal nature of this light curve is likely to be due to starspots on the surface. The model published by \citet{2017NewA...52....8K} included a cool spot on the secondary with a temperature factor of 0.9 and an angular size of 20\d. 

The American Association of Variable Star Observers (AAVSO) followed \targ\ during its 2016 outburst and also contributed to the \textit{HST} observations by monitoring the target immediately before and during the observations to ensure that an outburst did not saturate and potentially damage the instrument.

%\begin{figure}
%\centering
%\includegraphics[width=\columnwidth]{plots/long-ASAS-SN-crop.pdf}
%\caption{ASAS-SN photometry for V1460 Her from JD 2456231.6923 to JD 2458758.60393. The green triangles are V-band photometry and the blue circles g-band.  Evidence %for an outburst appears at JD 2457535. }\eb{=2016 May 26, so a month after the outburst we observed?} 
%\label{fig:long_asas-sn}
%\end{figure}

\subsection{WHT+ISIS spectroscopy}
\label{spectroscopy}
We obtained spectroscopy for a total of six nights with the target in a quiescent state during all of the observations. A log of these observations is shown in Table~\ref{tab:observations}. All of the observations were taken with the ISIS spectrograph mounted on the William Herschel Telescope (WHT) in La Palma, Spain. The gratings we used were the R1200B (centred on 4600\,\AA) with a spectral resolution of 0.85\,\AA\ and the R1200R (centred on 6400\,\AA) with a spectral resolution of 0.75\,\AA. The slit width of ISIS was set to 1\,arcsec for all observations. In total 618 science spectra were obtained, of which three were discarded due to poor signal to noise ratio which was probably caused by variability in the transparency of the sky. On most nights, flux standards and radial velocity standards were taken at the start and end of the observations and, on the last night, we also took spectra of three K-type (K0V, K5V and K7V) stars to help to characterise the secondary. For the observations on the last two nights, the nearby star 2MASS J16211745+4413386 (40\,arcsec north of V1460 Her; Gaia $G = 16.39$) was placed further along the slit. Although we used the extracted spectra of this star as a reference to monitor for changes in the transparency of the atmosphere during these two nights, since we did not have this coverage throughout the whole data set, it was not used systematically as a flux correction aid for our analysis.  

All of the spectra were optimally extracted and reduced using the {\sc pamela} and {\sc molly} reduction software, \citep{MarshPamela}. In order to perform flux calibration we fitted a spline to the continua of spectrophotometric standard stars taken on the same night as the observations and used their published flux as a reference, see e.g. \citet{Marsh1990}. For wavelength calibration, Cu+Ne and Cu+Ar arcs were observed several times over the course of the observations. For the red arm, we were then able to further improve the wavelength calibration by computing the shifts of sky emission features and applying them to the science spectra. As a final step in wavelength calibration, we observed at least one radial velocity standard each night and used this observation as a calibration for converting wavelengths to radial velocities.

%\begin{table}
%  \caption{Spectroscopy used in this study. The instrument used was ISIS mounted on the William Herschel Telescope.}
%  \label{tab:spectroscopy}
%  \begin{tabular}{l l l l}
%  \hline
%  Date & Gratings & $T_{\mbox{exp}}$   & Total time \\
%       &          & (s)                & (min) \\
%  \hline
%    2016/07/11  & R1200B(4601) R1200R(6400) & 54x150\,s & 135  \\
%    2016/08/16  & R1200B(4601) R1200R(6400) & 84x150\,s & 210  \\
%    2017/06/20  & R1200B(4601) R1200R(6400) & 153x150\,s & 382 \\
%    2017/06/21  & R1200B(4601) R1200R(6400) & 143x150\,s & 357 \\
%    2017/07/03  & R1200B(4601) R1200R(6400) & 107x150\,s & 267 \\
%    2017/07/04  & R1200B(4601) R1200R(6400) & 77x150\,s & 192  \\
%  \hline
%  \end{tabular}
%\end{table}

\subsection{\textit{HST}+COS spectroscopy}
\label{HSTobservations}
%\bc{when we describe the COS spectrum ... do we mention somewhere that this is probably the white dwarf photosphere absorbed by the ``iron curtain''?}
On 2017 March 02 (JD 2457815) we obtained \textit{HST} far-ultraviolet spectroscopy using the Cosmic Origins Spectrograph  \citep[COS,][]{2012ApJ...744...60G}. The data were collected over three consecutive primary spacecraft orbits (programme ID 14893), for a total of 8306\,s of on-source exposure. We observed the target using the Primary Science Aperture, the TIME-TAG mode and the far-ultraviolet grating G140L, centred at 1105\AA. The observations were dithered along the spectral direction on the detector using all four FP-POS settings, mitigating fixed pattern noise and detector artefacts. This delivered a spectrum covering the wavelength range of 1026 -- 2278\,\AA\, at a spectral resolution of $R\simeq3000$.
%In an effort to characterise the temperature of the White Dwarf, \textit{Hubble Space Telescope (HST)} observations were taken with the COS/FUV camera and the G140L filter with the central wavelength set at 1280\,\AA. The total exposure time was 8306\,s. 

The \textit{HST} spectra were also used to derive ultraviolet  photometry. The TIME-TAG observing mode of COS allows the recording of the time of arrival, the position on the detector and the pulse height of each detected event every 32\,ms. It is therefore possible to reconstruct a 2D time-resolved image of the detector, where the dispersion direction runs along one axis and the spatial direction along the other, from which a light curve of the system can be extracted. Following the method described in \cite{Pala+2019}, we identified three rectangular regions on this image: a central one enclosing the object spectrum and two regions to measure the background, one above and one below the spectrum. To extract a light curve of the white dwarf, we used the wavelength range $1175\ < \lambda < 1880$\,\AA\ and masked the geocoronal emission lines from Ly$\alpha$ ($1200 < \lambda < 1224\,$\AA) and \ion{O}{i} ($1290\ < \lambda < 1311\,$\AA), as well as the most prominent emission features from the accretion disc:
\ion{N}{v} ($1231  < \lambda < 1250\,$\AA), 
\ion{Si}{iv} ($1380 < \lambda < 1415\,$\AA), 
\ion{N}{i} ($1481 < \lambda < 14943\,$\AA), 
\ion{He}{ii} ($1635 < \lambda < 1646\,$\AA). We binned the data into five second bins and counted the number of events in each of those corresponding to the stellar spectrum. Once corrected for the number of counts from the background, these data provide the light curve of the target in counts per second. We discuss the spectrum and the derived photometry in sections \ref{sec:hstspectrum} and \ref{sec:hstphotometry}.

\section{Results}
\subsection{Photometric behaviour in outburst}
\begin{figure*}
\centering
\includegraphics[width=\textwidth]{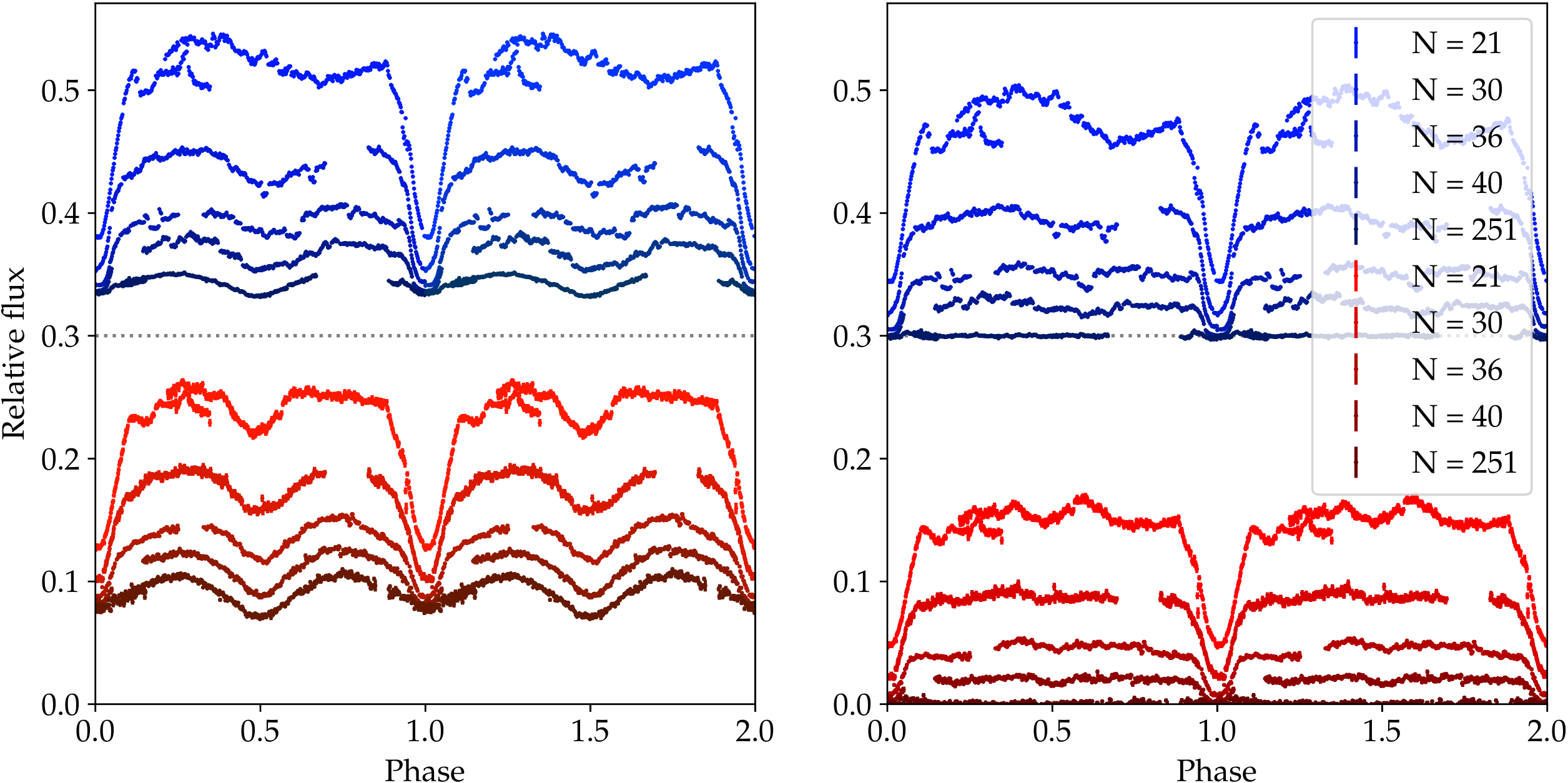}
\caption{The W1m photometry for \targ. phase folded according to equation \ref{eq:ephemeris}. The upper panels show the data for the BG40 (blue) filter (offset by 0.3) and the lower plots show the Z-band (red) filter. The relative flux is computed from the same comparison star each night. No offset has been introduced to data on subsequent nights, so the decrease in relative flux is due to the target fading from outburst to quiescence. The value $N$ in the plot legend refers to the approximate number of orbits since the start of the outburst (as calculated from the information given in \citet{2016ATel.9113....1T}, JD 2457541.113). The right hand panel shows the relative flux \textit{after} subtracting the quiescent light curve (N=251) from the originals.} \label{fig:stacked}
\end{figure*}

\begin{figure}
\centering
\includegraphics[width=\columnwidth]{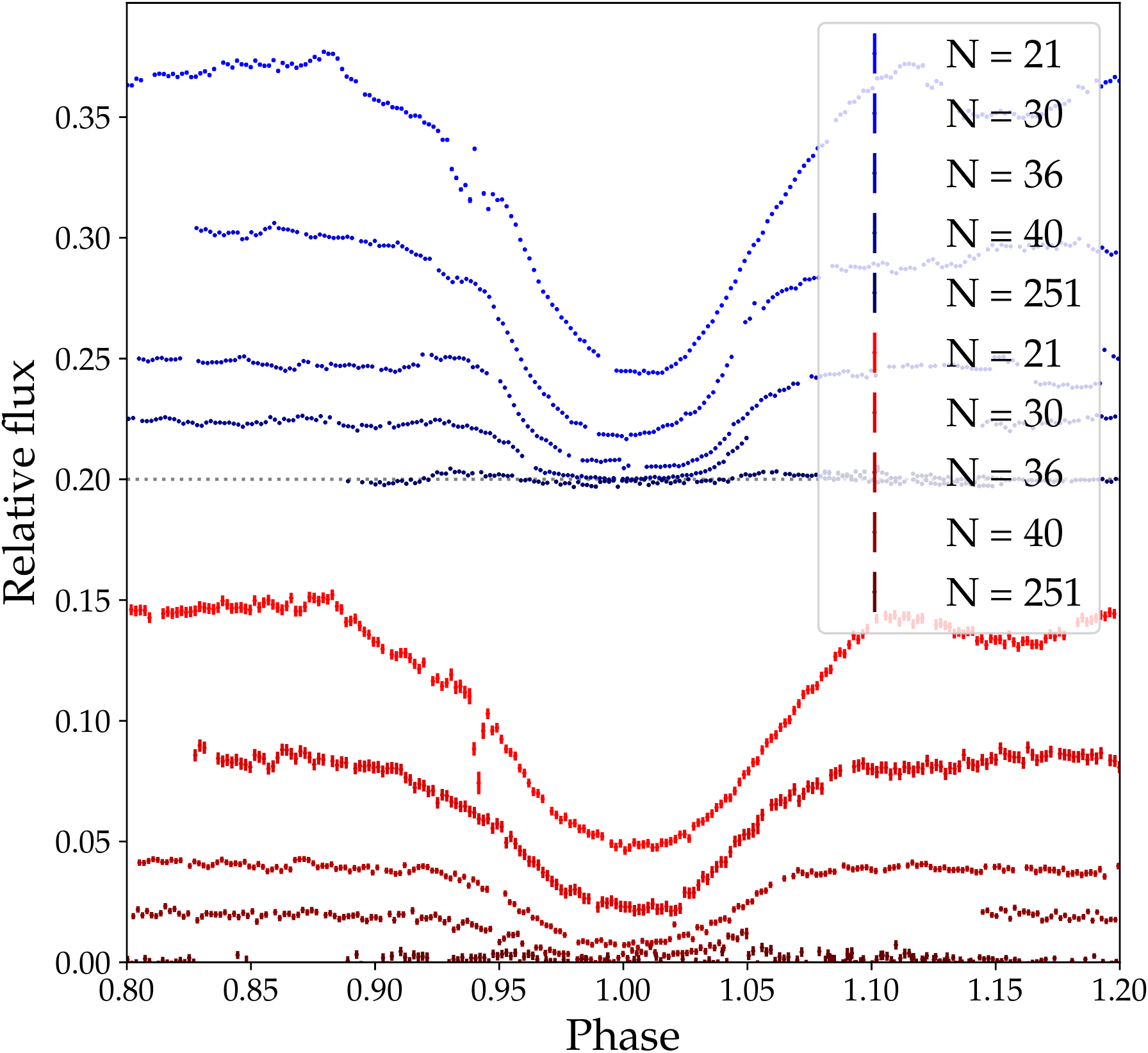}
\caption{Similar to Fig.~\ref{fig:stacked}, but in this figure we limit the phase to show primary eclipse in more detail. The blue data have been offset by 0.2.}
\label{fig:eclipse}
\end{figure}

The W1m photometry in Figs.~\ref{fig:stacked} and \ref{fig:eclipse} shows the gradual progression of the target from outburst to quiescence. For all of the observations, the same comparison was used for calibration. This means that the relative flux is calibrated against the same baseline. Since no offset has been applied to any of the light curves shown, the evolution of overall brightness of the system from outburst to quiescence is directly represented in the figures. The timescale of outburst to quiescence is demonstrated in Fig. \ref{fig:outbursts} at around 15 days. Our final night's data was taken some 52 days after the recorded outburst so we can be sure that the system had fully resumed quiescent behaviour. In order to better represent the contribution of the disc in our figures, the quiescent light curve was fitted with a sinusoid (and some harmonics) and this was subtracted from each night's data.  

\begin{figure}
\centering
\includegraphics[width=\columnwidth]{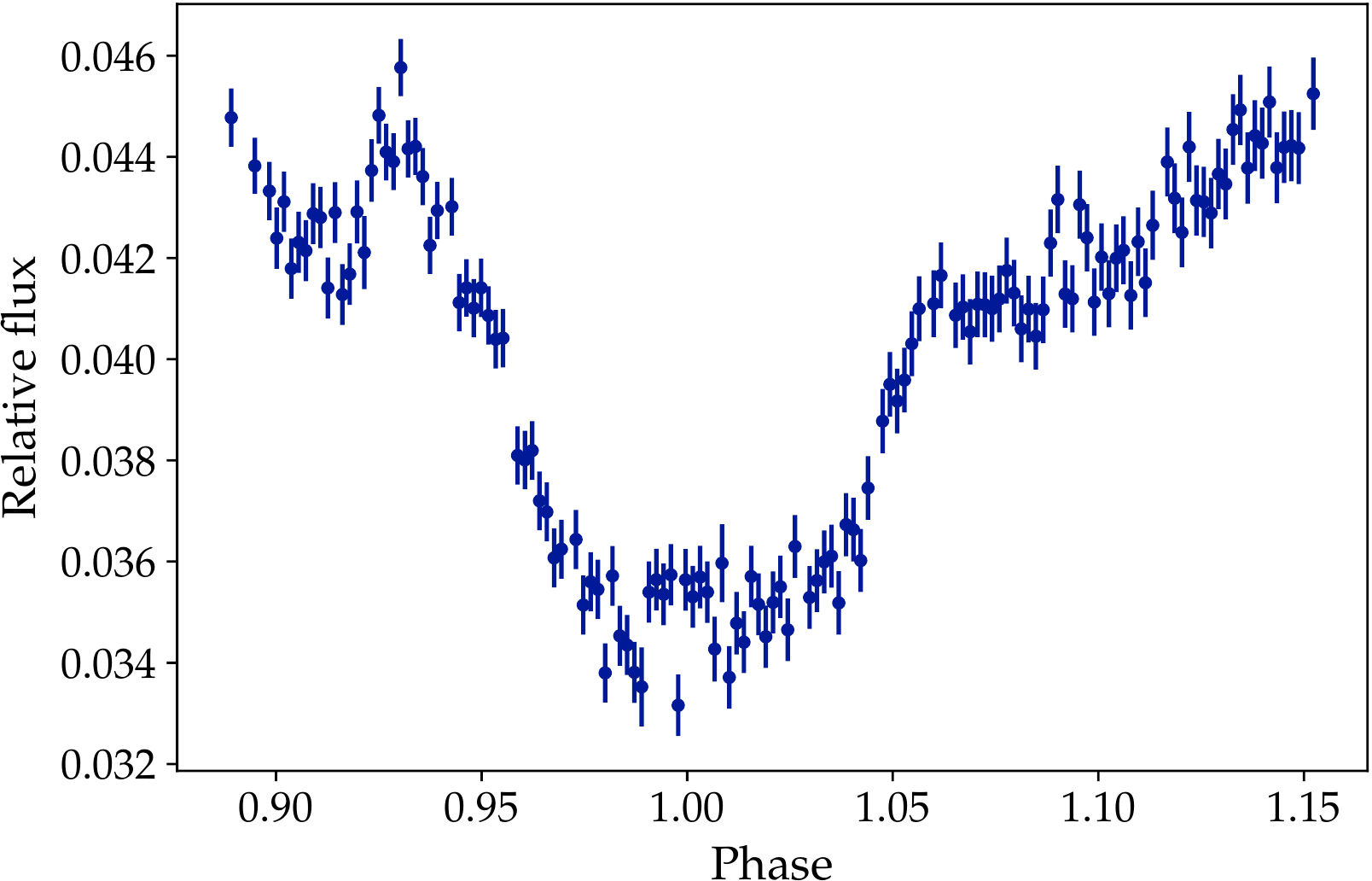}
\caption{A closer view of the primary eclipse taken on the last night of our W1m observations, occuring approximately 251 orbits since the 2016 June outburst with \targ\ back in a quiescent state. The profile of the eclipse and the rate of transition ($\simeq 7.5$\,min) during ingress and egress indicate that we are seeing the eclipse of a disc rather than the white dwarf itself.}
\label{fig:quiescenteclipse}
\end{figure}

Our outburst data confirm the behaviour reported by \citet{Zola2017} and show dominance by an accretion disc during outburst, transitioning to the quiescent light curve which is dominated by the tidally-distorted secondary star. We have one observation of the primary eclipse taken during quiescence, Fig. \ref{fig:quiescenteclipse}, which shows a U-shaped profile and a fairly long transition during ingress and egress. Based on this evidence and the supporting data from spectroscopy, we propose that the disc, although dimmed during quiescence, is still present and we cannot discern the white dwarf. 

\subsection{Ephemeris}
\begin{figure}
\centering
\includegraphics[width=\columnwidth]{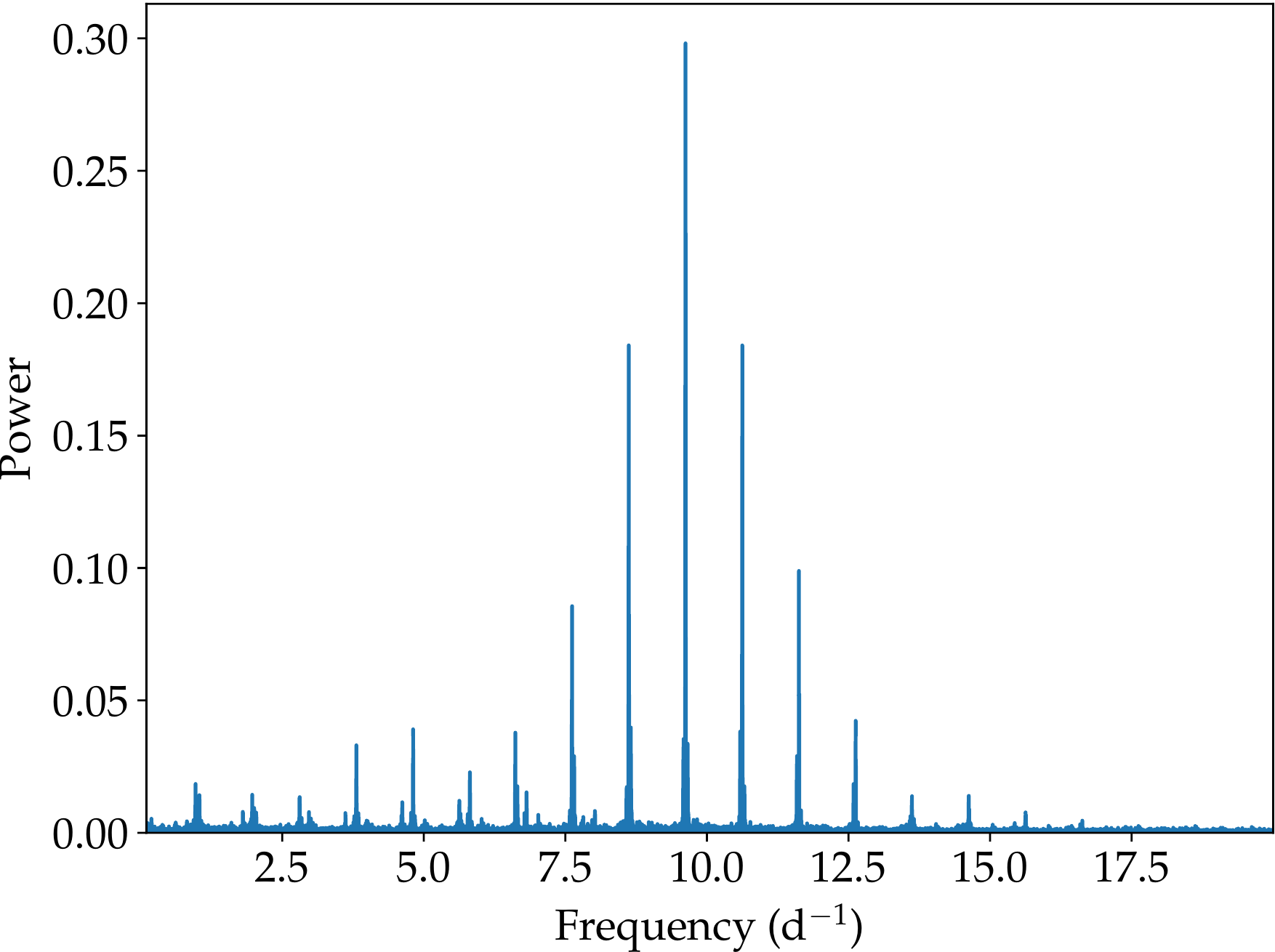}
\caption{Lomb Scargle analysis of the survey photometry contained in the ASAS-SN, SuperWASP and CRTS archives. The peak of the power lies at  9.622231\,d$^{-1}$, but, since the lightcurve is dominated by the ellipsoidal variation of the secondary, we halved that frequency to give the orbital period of the system. This was then used as a starting point to calculate the ephemeris shown in equation \ref{eq:ephemeris}.}
\label{fig:lombscargle}
\end{figure}

Since there is a long baseline of over 14 years of published survey photometry for \targ, all of these data were combined and used to find its orbital period. We used the SuperWASP, CRTS and ASAS-SN observations and computed a Lomb-Scargle periodogram \citep{1982ApJ...263..835S} to determine an initial estimate for the period as is shown in Fig. \ref{fig:lombscargle}. This was then further refined by fitting a sinusoid plus four harmonics to all of the data. Once we had the period, the WHT spectroscopy was folded on this period and as zero point we adopted the time of blue-to-red crossing of the radial velocity of the secondary (\ref{sec:orbital}). Our ephemeris for \targ\ is 
\begin{equation}\mathrm{BJD(TDB)} = 2457760.469657(6) + E\times0.20785223(3),\label{eq:ephemeris}\end{equation}
and gives the times of primary eclipse, when the donor star is closest to Earth. Times are referenced to the Barycentric Julian Date (BJD) in Barycentric Dynamical Time (TDB).

\subsection{Spectral type of the secondary}
\begin{figure}
\centering
\includegraphics[width=\columnwidth]{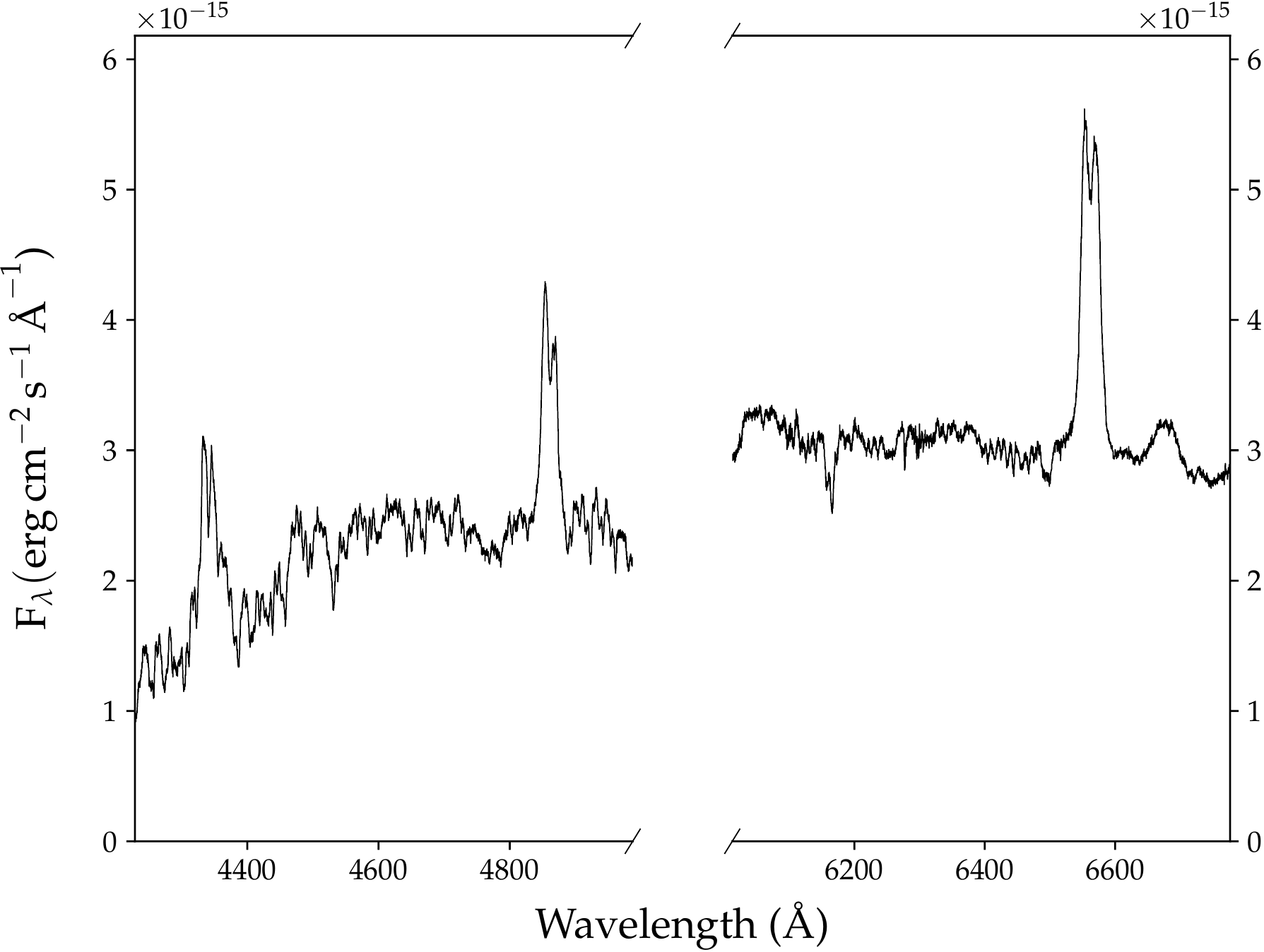}
\caption{The overall average spectrum of the ISIS red and blue arms. The average was computed with respect to a heliocentric rest frame. }
\label{fig:combined_spectra}
\end{figure}

Fig.~\ref{fig:combined_spectra} shows an average spectrum across all phases of the orbit.  This shows a combination of absorption lines from the secondary and double-peaked Balmer emission from the disc. Although we detect weak \ion{He}{i} emission at 6678\,\AA, there is no detection of \ion{He}{ii} emission. 

The individual target spectra show many absorption features from the secondary star which move in approximate anti-phase to the emission lines. Starting with a hand-picked spectrum as a reference, we masked out a region of the continuum that was free of emission features (6095 -- 6426\,\AA) and performed a cross-correlation of each target spectrum with this reference. Before cross correlation, all of the spectra were binned into the same wavelength scale and a spline fit to the continuum was subtracted.  The cross correlation produced radial velocities that we fit with a sinusoid in order to derive an initial estimate for the secondary's semi-amplitude, $K_2$, which we subsequently refined once we had a good template match of the secondary, see Section \ref{sec:orbital}. The ephemeris and this $K_2$ estimate was then used to generate an average spectrum as computed in the rest frame of the secondary. 

Since we had anecdotal evidence that the secondary had a K-type spectral classification, as noted by \citet{2016ATel.9141....1T}, we took three additional spectra of known K-type stars on our last night of WHT observations, namely HD~124752 (K0V), HD~122120 (K5V) and HD~234078 (K7V). For our analysis, we supplemented these by adding eight more K-type spectra taken at the neighbouring Isaac Newton Telescope as templates for a similar study of AE~Aqr \citep{10.1093/mnras/282.1.182} and sent to us in a private communication by Jorge Casares. The complete list of the templates we used is shown in Table \ref{tab:vsini_analysis}. 

\begin{table}
  \centering
  \caption{K-type templates matched to the spectrum of the secondary to determine a value of $v \sin i$. The $\alpha$ value is the fractional contribution of the template applied to best match our target spectrum. Asterisks mark the templates that were taken during our observations at the WHT.  The other templates were collected from archived data. All computations of $\chi^2$ had 1500 degrees of freedom.}
  \label{tab:vsini_analysis}
  \begin{tabular}{l l c c}
  \hline
  Name and spectral type & $v \sin i\,(\kms)$ & $\alpha$ & $\chi^2$ \\
  \hline
    HD124752 (K0V) & 110.4 $\pm$ 0.5 & 1.26 & 7628 \\
    HR8857 (K0V) & 109.8 $\pm$ 0.6 & 1.23 & 9282 \\
    HD184467 (K2V) & 110.6 $\pm$ 0.6 & 1.39 & 6536 \\
    HD154712 A (K4V) & 110.2 $\pm$ 0.6 & 1.05 & 4529 \\
    HD29697 (K4V) & 108.8 $\pm$ 0.7 & 1.01 & 4171 \\
    61Cyg A (K5V) & 109.8 $\pm$ 0.6 & 0.99 & 4491 \\
    HD122120 (K5V) & 109.8 $\pm$ 0.6 & 0.85 & 2997 \\
    HD234078 (K7V) & 109.4 $\pm$ 0.6 & 0.81 & 5439 \\
    61Cyg B (K7V) & 108.7 $\pm$ 0.7 & 0.95 & 7555 \\
    HD154712 B (K8V) & 111.0 $\pm$ 0.7 & 0.89 & 9256 \\
  \hline
  \end{tabular}
\end{table}

\begin{figure}
\centering
\includegraphics[width=\columnwidth]{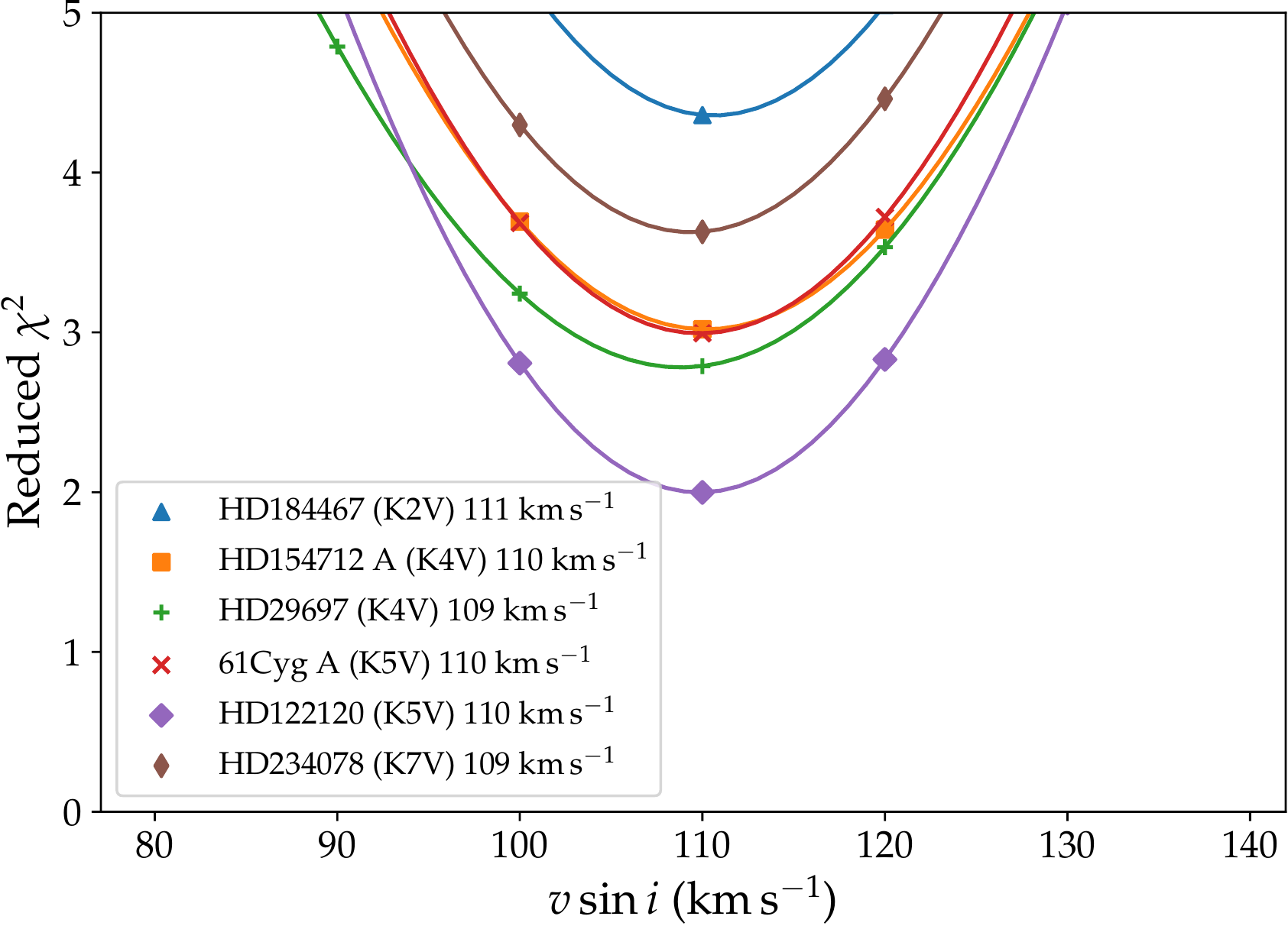}
\caption{The best matching K-type templates as a function of varying values chosen for $v \sin i$. $\chi^2$ values were computed after performing an optimal subtraction of an artificially broadened version of the template spectrum from the target spectrum. Some templates with poorer matches have been omitted from this figure, but are included in Table \ref{tab:vsini_analysis}.}
\label{fig:optsubs_chisq}
\end{figure}

\begin{figure}
\centering
\includegraphics[width=\columnwidth]{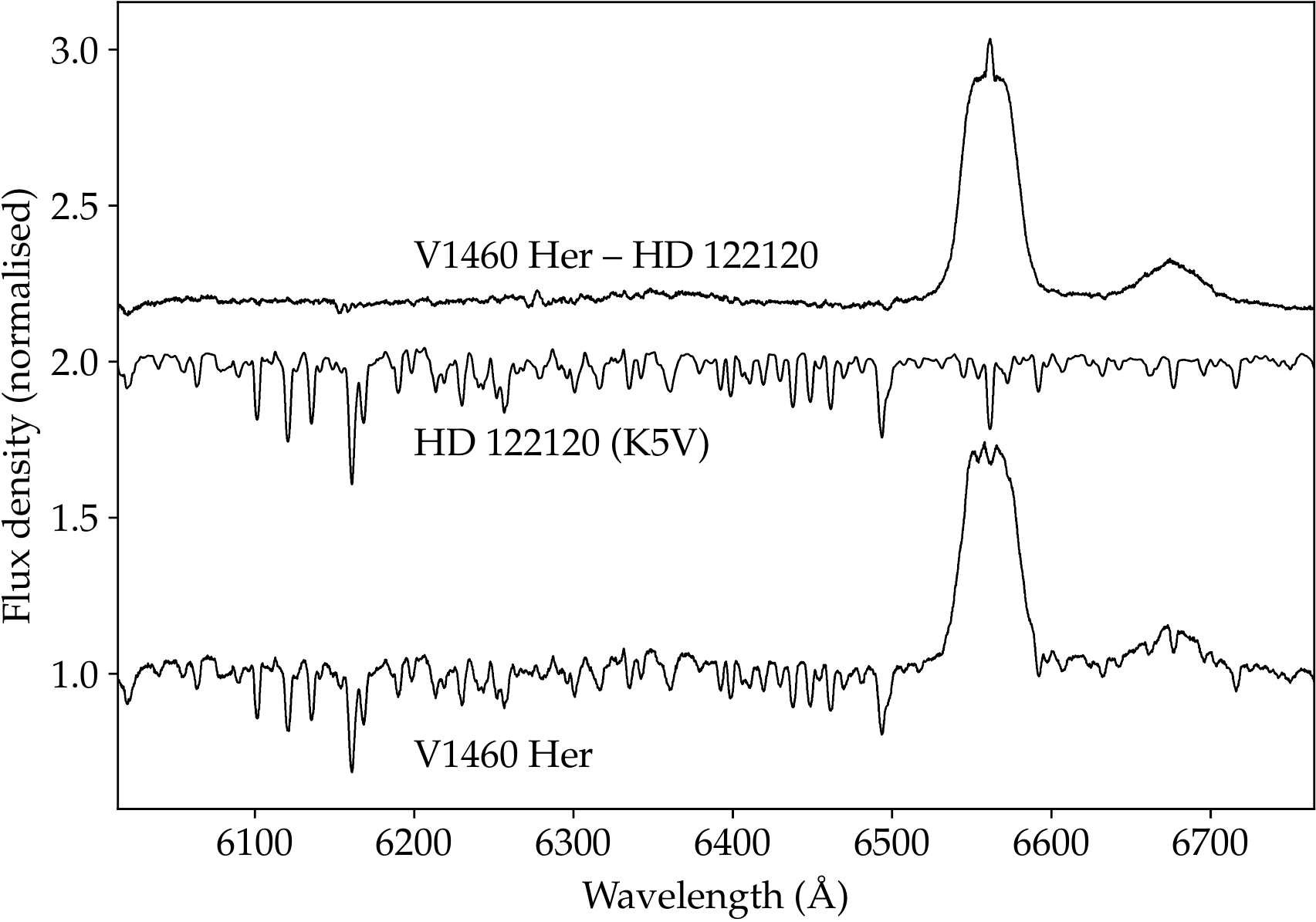}
\caption{The lower spectrum is the average of the target once shifted into the rest frame of the secondary. The middle spectrum is that of HD 122120 (K5V) after application of rotational broadening with the value $v \sin i$ = 109.8 km\,s$^{-1}$. The upper spectrum is the result of subtracting the two.}
\label{fig:optsub}
\end{figure}

By artificially introducing rotational broadening to these spectra and then comparing them to our target spectra we were able to find the value of $v \sin i$\ that gave us the best match to the donor. This was done by stepping through values of $v \sin i$\ from 80 to 140\,$\kms$\ in increments on 1\,$\kms$\ and computing an optimal subtraction with a variable scaling constant $\alpha$\ (or fractional contribution) applied to each template. These results are summarized in Table \ref{tab:vsini_analysis} and the $\chi^2$ versus $v \sin i$\ relationship is shown in Fig. \ref{fig:optsubs_chisq}. The spectrum of HD~122120 (K5V) with a rotational broadening value of $v \sin i = 109.8 \pm 0.6\,\kms$ gave us our best subtraction. The minimum value was computed by fitting a simple quadratic polynomial to the three values closest to the minimum in $\chi^2$ and the error was estimated by computing the range around this minimum that increased the value of $\chi^2$ by one. In Fig.~\ref{fig:optsub} we show the averaged target spectrum (in the rest frame of the secondary), the best-fit rotationally broadened spectrum of HD~122120 and the result of the subtraction of the two. 

\begin{figure}
\centering
\includegraphics[width=\columnwidth]{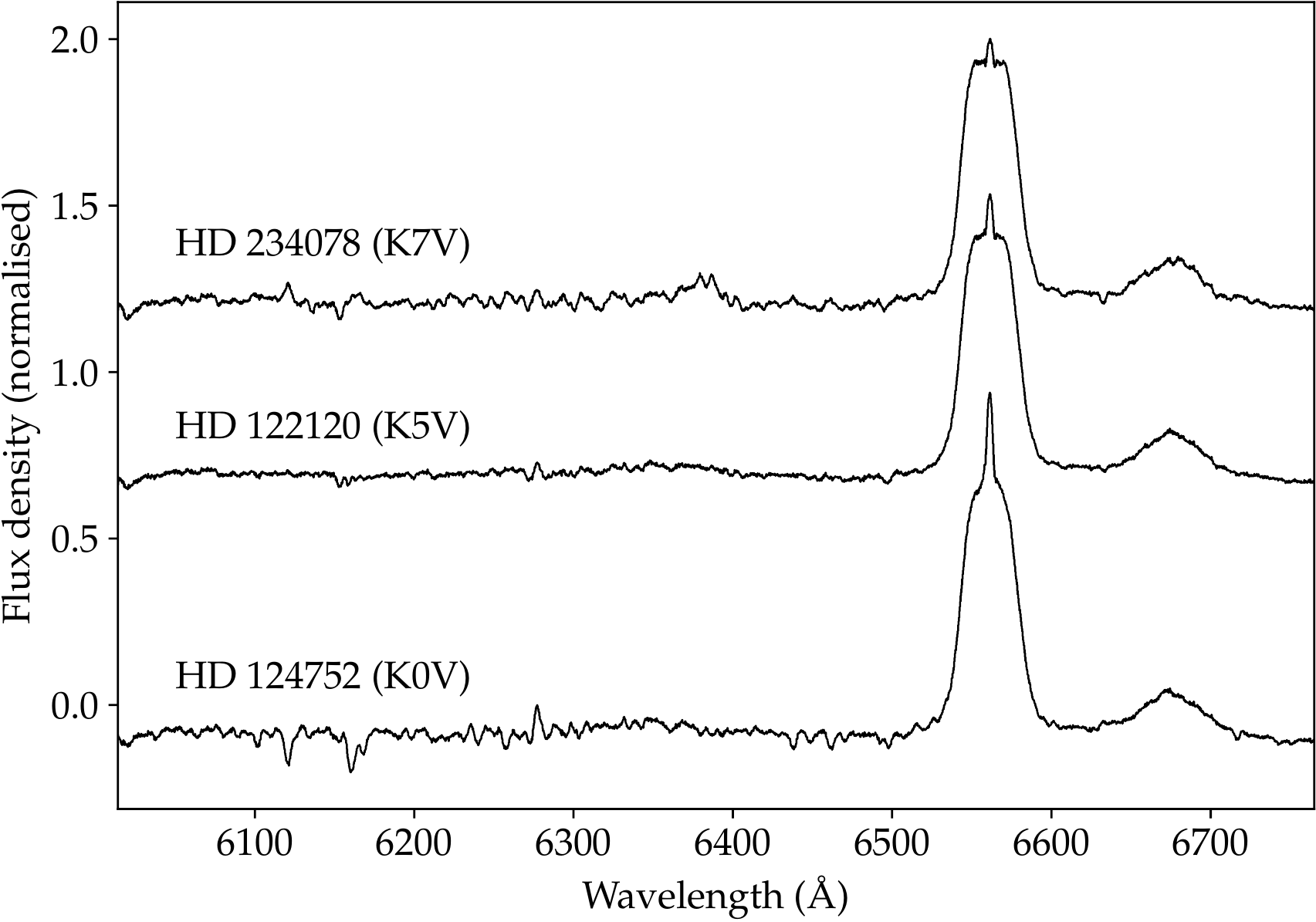}
\caption{The results of subtracting a rotationally broadened version of the three template spectra from the time averaged spectrum in the rest frame of the secondary. Our best results were achieved for a spectral type of K5V. The flux density is normalised and an offset of 0.5 units has been applied to each subsequent spectrum.}
\label{fig:templates}
\end{figure}

\subsection{Orbital radial velocities}
\label{sec:orbital}
\begin{figure}
\centering
\includegraphics[width=\columnwidth]{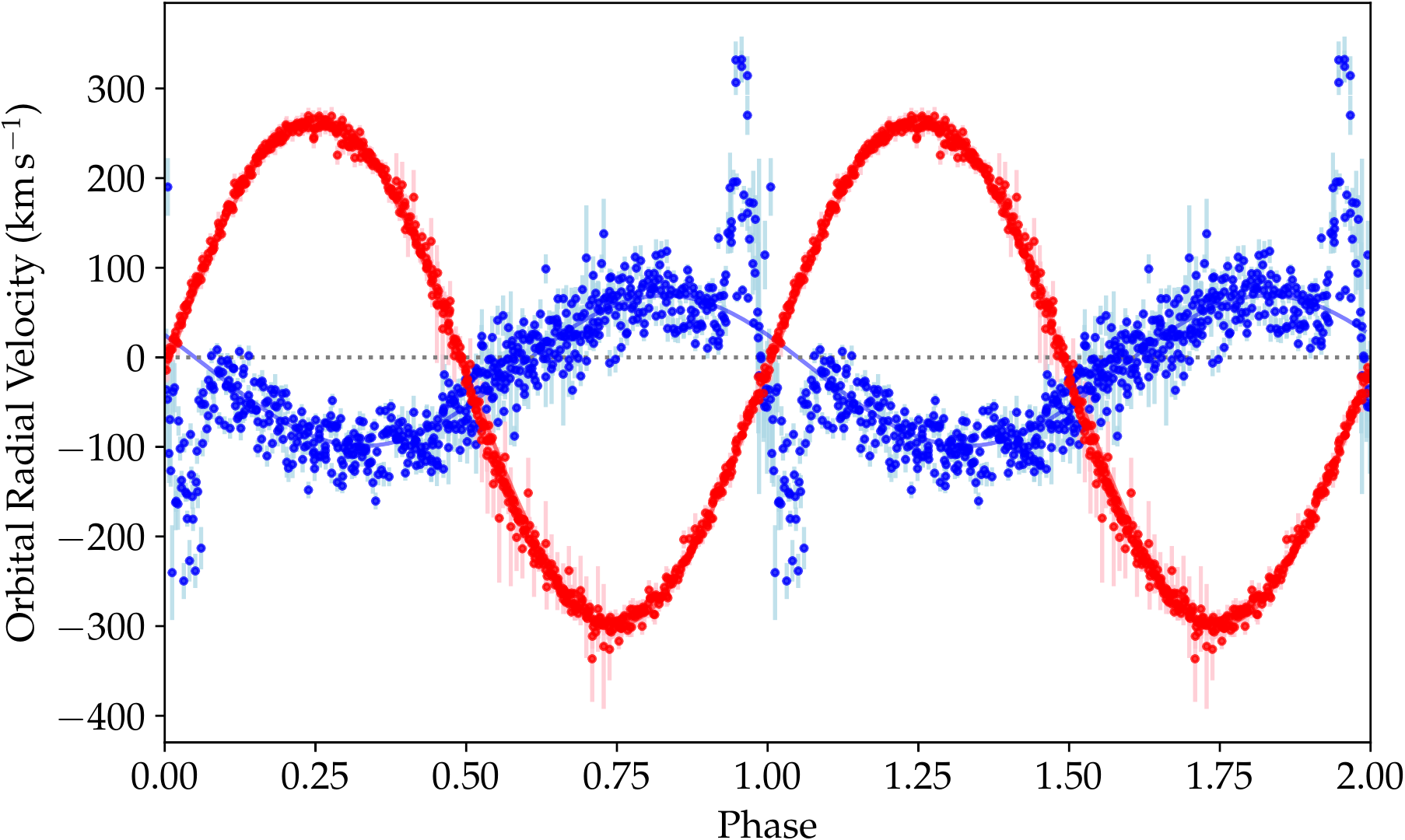}
\caption{Sine fits to the radial velocities of the absorption and emission features in the spectra of \targ. The absorption velocities were computed by cross correlation with HD~122120. Those for the emission were computed with a Young and Schneider analysis of the H$\alpha$ feature. The emission radial velocity profile exhibits a rotational disturbance near the primary eclipse as first blue-shifted and then red-shifted gas is eclipsed in turn, described in Section \ref{sec:rotational-disturbance}. For the calculation of the orbital parameters listed in Table \ref{tab:ephem}, the points between phases 0.90 and 1.10 were excluded.}
\label{fig:rvs}
\end{figure}

We next measured the radial velocity amplitude, $K_2$, and systemic velocity, $\gamma_2$, of the donor star by cross-correlating the target spectra with the rotationally broadened K5 template (HD 122120). This template star has a well defined radial velocity of $-57.3 \pm 0.03\,\kms$\ \citep{RVstandards}. The rotationally broadened version of its spectrum was cross correlated with the science spectra for our target. Once again, we limited the analysis to a region of the spectrum that is dominated by flux from the secondary and contains no emission lines, 6095 -- 6426\,\AA. The cross correlation process resulted in 615 radial velocities. These were fitted with a sinusoid of the form $v(t) = \gamma_2 + K_2 \sin({2\pi}(t - T_0) / P)$, see Fig.~\ref{fig:rvs}. Here, the value of the orbital period, $P$, was fixed to the value given in equation \ref{eq:ephemeris}, derived from the long baseline photometry. $T_0$ is derived from the zero-crossing point of the radial velocities going from the negative to the positive regime, representing the phase of the orbit when the secondary is closest to the Earth. The derived parameters are listed in Table~\ref{tab:ephem}. 

The strongest emission feature in the optical spectra is the Balmer H$\alpha$ line. We measured the radial velocity in the line wings using a technique described in \citet{1980ApJ...238..946S}, involving cross-correlation of emission line profiles with two Gaussian profiles separated by a fixed value. In order to determine the optimal separation of the Gaussian profiles, we performed an exhaustive search by stepping through values of this constant, starting at 800 and going up to 2400\,$\kms$\, in increments of 10\,$\kms$. For each of these chosen separations, we examined the quality of the correlation by measuring the mean error in the derived radial velocity values following a technique described by Shafter 1983. Our chosen value for the optimal separation of the Guassian profiles was 1800\,$\kms$.  These radial velocities were fitted with a sine function that had the period fixed to the one given in equation \ref{eq:ephemeris}. The systematic radial velocity, K$_1$ semi-amplitude and phase offset were derived from the fit and are listed in Table \ref{tab:ephem}.  

The zero-crossing points in the radial velocity measurements differ by 0.059 in phase which corresponds to 21\d. \citet{Stover1981} noticed this effect in several systems, and explained it by an area of enhanced emission associated with region where the accretion stream impacts the disc. \citet{1987MNRAS.225..551M} suggested that for Z~Cha this could be explained by a combination of the distortion of the velocity field in the inner disc and variations in the strength of the emission over the disc. We are interpreting this phase shift similarly and proposing that the emission radial velocities are affected by the asymmetry of the bright spot region. 

\begin{figure}
\centering
\includegraphics[width=\columnwidth]{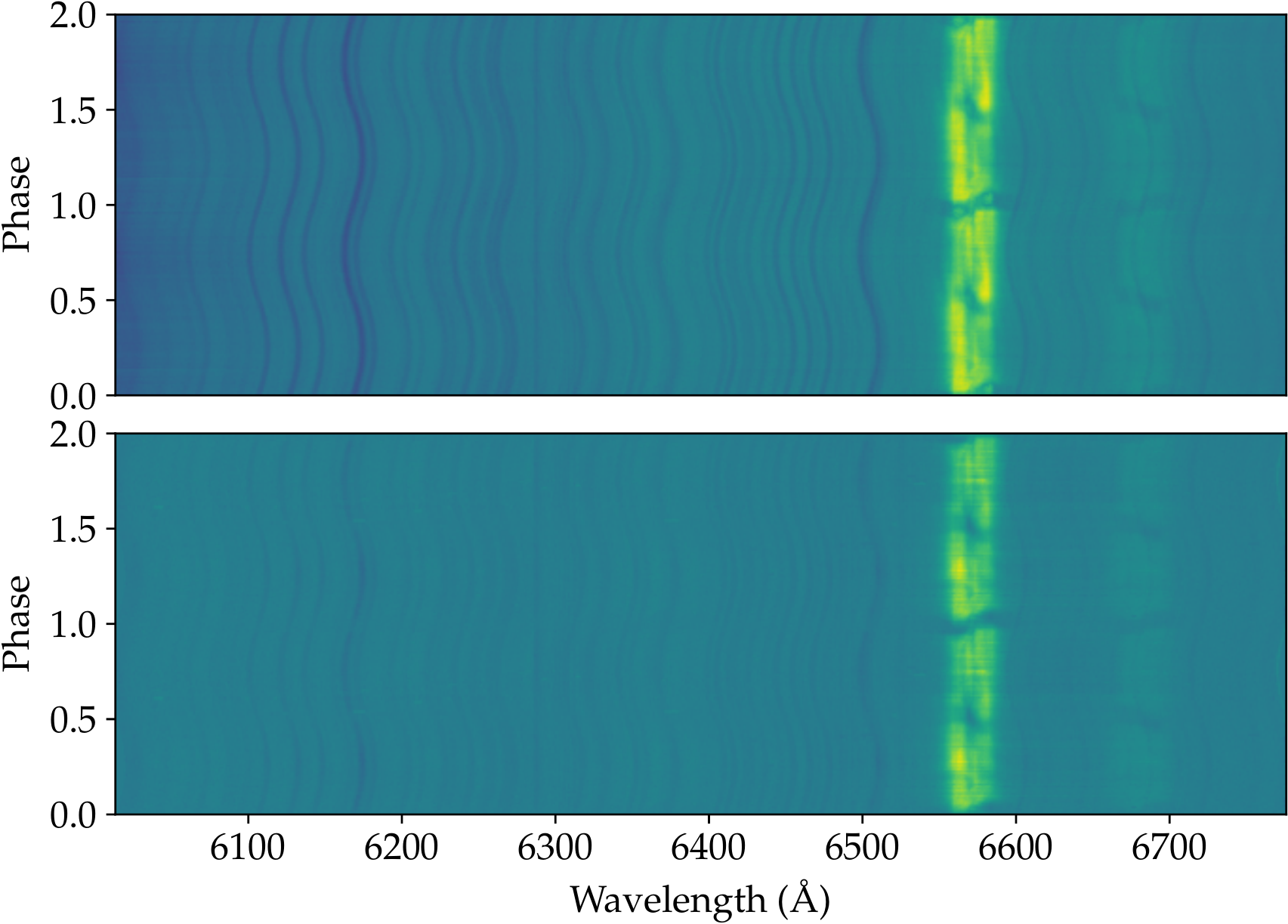}
\caption{Trails of the spectra taken in the red arm of ISIS. The upper trail shows the original data in the heliocentric rest frame. The lower trail shows the result after subtracting a rotationally broadened K5 star template to remove the contribution from the secondary. The emission lines on the right are H$\alpha$ and HeI~6678.}
\label{fig:redtrails}
\end{figure}

For each of the 615 science spectra, the rotationally broadened spectrum of HD 122120 was shifted to match the correct radial velocity for the phase in the orbit and then subtracted. Since the secondary's contribution to the overall flux is highly dependent on the phase of the orbit (see Fig.~\ref{fig:bin_swasp}), each shifted template spectrum was multiplied by a phase dependant constant to reflect its proportional contribution. The resulting trailed spectra diagrams are shown in Fig.~\ref{fig:redtrails}. 

\subsection{Masses of the components}
Assuming co-rotation, the equatorial speed of a Roche-lobe filling star is related to the orbital speed of its centre of mass by a function of the mass ratio $q$ alone. Using  \citet{Paczynski1971}'s equation for the radius of the secondary star's Roche lobe leads to
\begin{equation}
	\frac{v \sin i}{K_2} = 0.46[(1+q)^2 q ]^{1/3}, 
\end{equation} 
 for 0 < $q$ < 0.8 \citep{1988ApJ...324..411W}.
 
Using our values for $K_2$ and assuming the value of $v \sin i$ derived in the previous section, we calculate 
$q = 0.337 \pm 0.006$. For this mass ratio, the corresponding projected velocity semi-amplitude of the primary,  K$_1$\, ($v \sin i$)\,, would be 95.2 $\pm 1.7 \,\kms$ which is higher than the 84.2 $\pm 1.5 \,\kms$ value measured from the H$\alpha$ emission. While these two figures are not in agreement, it should be noted that they are measured from different physical objects within the system, the inner disc and the rotational broadening of the secondary.  

Examining both the W1m and \textit{HST} photometry of the primary eclipse, it is possible to estimate the duration of the white dwarf's eclipse, $\Delta \phi$, the mean phase full-width of eclipse at half the out-of-eclipse intensity. Using the geometry of the Roche lobe and an estimate of the mass ratio, the inclination of the orbit can be deduced, as described by \citet{1992MNRAS.258..225D}. We derived $\Delta \phi$ by measuring the FWHM of the eclipse profile for each of the 4 nights of photometry taken during outburst on the W1m in both the red and blue filters and also from the ultraviolet photometry taken from the \textit{HST}. For the \textit{HST} data we had to assume the ingress was a symmetrical analogue of the egress. These measurements gave a value of $\Delta\phi = 0.094 \pm 0.009$, which leads to an inclination between 84\d to 90\d, or $\sin i \sim 0.997$.

The mass function for a compact binary system is given by,
\begin{equation}
 	\frac{(M_1 \sin i )^3}{(M_1+M_2)^2} = \frac{P_\mathrm{orb} K_2^3}{2\pi G} ,
\end{equation}
and assuming that $\sin i \sim 0.997$, we can derive a mass for each component of
\begin{eqnarray}
	M_1 &=& 0.869\pm 0.006 \, \msun,\\ M_2 &=& 0.295 \pm 0.004 \, \msun. 
\end{eqnarray}
 
Using equation 2 in \citet{Eggleton1983} for a Roche lobe filling secondary, we can derive a radius of $R = 0.430 \pm 0.002\,\rsun$ and a binary separation of $a = 1.478 \pm 0.003\,\rsun$. 
%Substituting the Roche lobe radius and our value of $q$ into \citet{Eggleton1983}'s equation 3, we can derive the average density of the secondary at  $\rho_2 = 4533 \pm 3\,\mathrm{kg}\,\mathrm{m}^{-3}$
%For comparison, a simple calculation of the average density of any object filling its Roche lobe is given by $\rho = 1.07\times10^5/P_\mathrm{orb}^2$. This can be derived by combining Kepler's law with \citet{Paczynski1971}'s equation for the radius of the secondary, $R_2 / a = 0.46224(\frac{q}{1+q})^{1/3}$. The resulting density for the secondary is $\rho$ = 4280\,{kg}\,m$^{-3}$ and therefore a radius of $R = 0.459 \pm 0.001\,\rsun$.
%\trmc{This is a little confusing: we are effectively performing the same calculation twice with two different approximations to the same underlying physics. The Eggleton value is more accurate so don't bother with the second Paczynski-based estimate. Should probably revise the $v \sin i$ stuff which also uses Paczynski.}

We can estimate the radius of the donor independently from \textit{Gaia} data since the donor dominates the total light.  The \textit{Gaia} DR2 data for \targ\ gives a parallax of $3.764\pm0.026$\,mas, corresponding to a distance of $265.7\pm1.8$\,pc. Using our estimate for the spectral type of the donor as a K5 star we chose a temperature of 4410\,K, as would be expected from a main sequence star of this type \citep{2013ApJS..208....9P}. We did not use the \textit{Gaia} published estimate of 4861\,K for \targ\ as this combines flux from all components of the binary and is subject to the variability of the target. We assumed a reddening of $E(B-V) = 0.02 \pm 0.02$ and calculated a line of sight extinction in the \textit{Gaia} G-band of $A_\mathrm{G} = 0.04 \pm 0.04$. Our reddening estimate was taken from \citet{Stilism2017}. Since our best template match (see Table \ref{tab:vsini_analysis}) had a secondary star fractional contribution of 0.85, we increased the value of the \textit{Gaia} magnitude by $-2.5 \log(0.85)$, taking the value from $G=15.03$ to $15.21$ and therefore more closely representing the flux from the secondary star only. Using equations 8.1, 8.2 and 8.6 of the \textit{Gaia} Data Release 2 documentation\footnote{https://gea.esac.esa.int/archive/documentation/GDR2/} we calculate an absolute magnitude $M_\textrm{G} = 8.05 \pm 0.18$ and $R_2 = 0.43 \pm 0.04 \,\rsun$.  The potential uncertainty in the determination of the donor's spectral type and hence its temperature is also a contributor to the uncertainty in this calculation. If we choose a K4 and a K6 type for comparison, our calculations of the radius become $R_2 = 0.39\,\rsun$ for a K4 donor and $R_2 = 0.46\,\rsun$ for a K6 donor.  The \textit{Gaia}-based estimate of the donor's radius is in agreement with the estimate of $0.43\,\rsun$ based on the measurement of $v \sin i$ as described in the previous paragraphs.  

\begin{table}
  \caption{Orbital and stellar parameters measured in this paper.  }
  \label{tab:ephem}
 \begin{tabular}{ l  l  }
    \hline
    $\mathrm{M}_1$   & 0.869 $\pm$ 0.006\,$\msun$ \\
    $\mathrm{M}_2$   & 0.295 $\pm$ 0.004\,$\msun$ \\
    K$_1$ ($v \sin i$)   & 95.2  $\pm$ 1.7 km\,s$^{-1}$   \\
    %K$_1$ (H$\alpha$)  & 84.1  $\pm$ 1.5 km\,s$^{-1}$   \\
    K$_1$ (H$\alpha$)  & 84.2  $\pm$ 1.5 km\,s$^{-1}$   \\
    $\mathrm{K_2}$   & 282.4 $\pm$ 0.4 km\,s$^{-1}$   \\
    $\gamma_1$       & -15.0 $\pm$ 1.2 km\,s$^{-1}$  \\
    %$\gamma_2$ (H$\alpha$)      & -15.3 $\pm$ 0.4 km\,s$^{-1}$  \\
    $\gamma_2$ (H$\alpha$)       & -18.1 $\pm$ 1.1 km\,s$^{-1}$  \\
    q                & 0.337 $\pm$ 0.006 \\
    a                & 1.478 $\pm$ 0.002\,$\rsun$ \\
    $\mathrm{R}_2$   & 0.430 $\pm$ 0.002\,$\rsun$ \\
    T$_0$            & 2457760.46887(6) d \\
    Period           & 0.20785223(3) d \\
    White dwarf spin period & 38.875 $\pm$ 0.005\,s \\
    Secondary spectral type & K5 \\
    \hline
  \end{tabular}
\end{table}

%  GAIA data
%		RA & Dec & Plx & Distance & Gmag & pmRA & pmDec \\
%		(\d) & (\d) & (mas) & (pc) & & (mas/yr) & (mas/yr)\\
%		245.32235560540(0.0194mas) & +44.21496799716(0.0243mas) & 3.764(0.026) & 265.7(1.8) & 15.035(0.012) & 6.602(0.037) &	-15.507(0.050)\\

\subsection{Spectral features}\label{sec:rotational-disturbance}
\begin{figure}
\centering
\includegraphics[width=6cm]{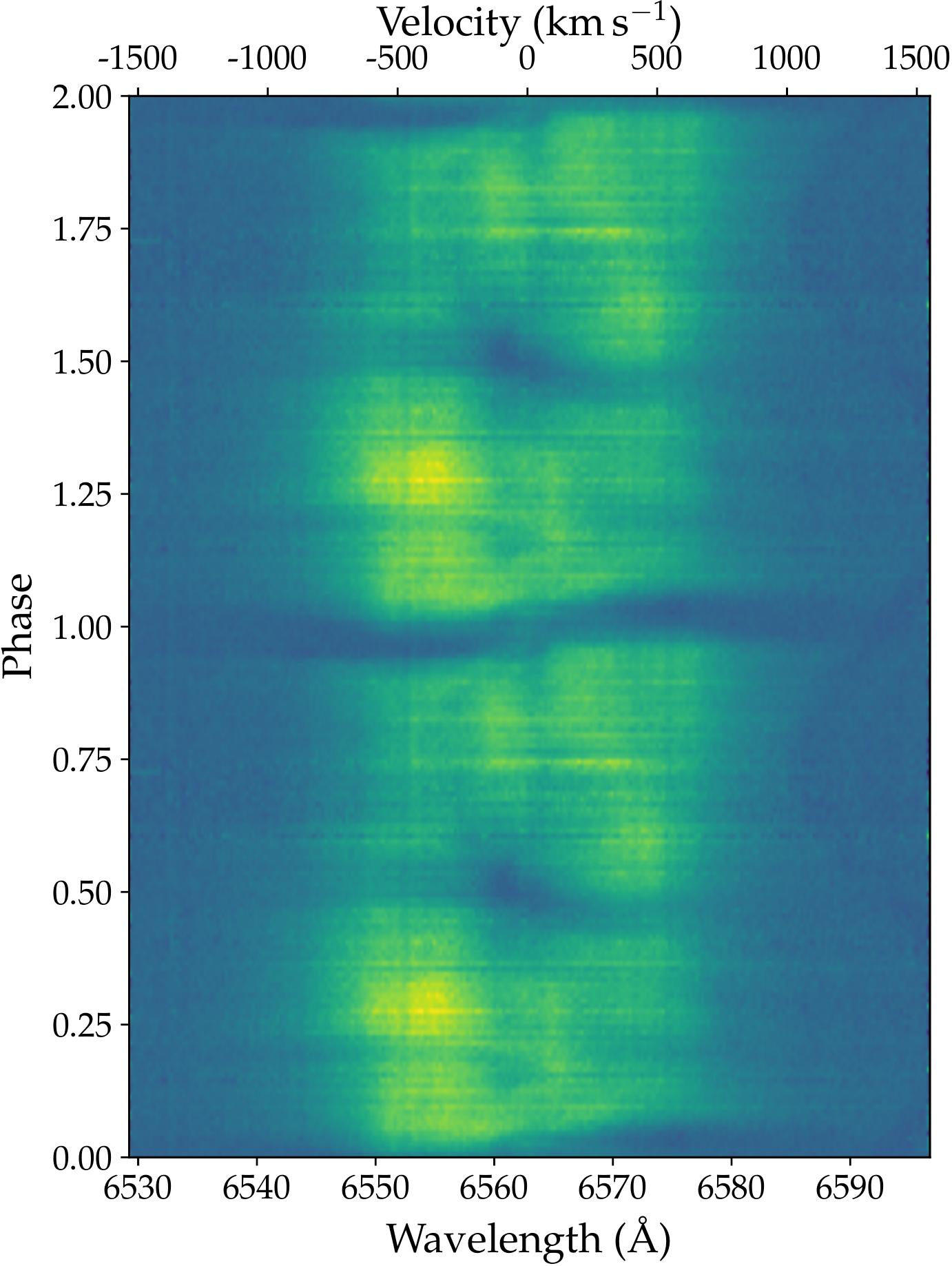}
\caption{The trailed spectra at H$\alpha$ for \targ. } 
\label{fig:halpha_subtracted}
\end{figure}

Our spectra, which were taken during quiescence, strongly support the presence of a disc, which is not clearly evident from photometry, except in outburst. The broad, double-peaked emission features are an important indication, and there is, in addition, evidence of a "rotational disturbance" (similar in nature to the Rossiter-McLaughlin effect) first seen in CVs by \citet{1959ApJ...130...99G}. The rotational disturbance is seen when the emission lines from a disc are eclipsed. If the disc orbits in a prograde direction within the binary, the approaching, blue-shifted side of the disc is first eclipsed, followed by the red-shifted, receding side. This can be clearly seen in the phases near to the eclipse in the trail of Fig.~\ref{fig:halpha_subtracted} and also in the emission line radial velocities of Fig.~\ref{fig:rvs}.

\targ\ shows an additional behaviour of particular interest. At around phase~0.5 there is an absorption feature in its trailed spectra diagram that is akin to an inverted rotational disturbance. We identify this with the "mirror eclipse" feature first found by \citet{LittlefairIPPeg}. They show that this feature results from the passage of the disc in front of a bright secondary star as light from the secondary star passes through the disc, and is partially absorbed by it. \citet{LittlefairIPPeg} discovered this phenomenon in K-band spectra of the dwarf nova IP~Peg which has a secondary star that dominates in the near infrared. \targ's bright donor makes this the first system where the mirror eclipse has been recognised at H$\alpha$. The key property needed to see a mirror eclipse is that there are lines-of-sight to the secondary that pass through parts of the disc which are optically thin in the continuum. Interestingly, the mirror eclipse in \targ\ does not seem to extend into the line wings and has no obvious influence upon the emission line velocities around orbital phase~0.5. This suggests that the inner disc is optically thick in the continuum, as is expected for accretion discs in CVs.

%The second feature of interest is highly unusual. During the eclipse of H$\alpha$ (Fig.~\ref{fig:halpha_subtracted}), the flux at each velocity at first decreases as expected, and with the phase offsets previously described, but then shows a local maximum at the centre of the eclipse, with the same phenomenon seen on both sides of the line, particularly around the line peaks. Only H$\alpha$ shows this local maximum. This is very odd behaviour, which we suspect could be associated with seeing the disc very close to edge on, combined with high optical depth in H$\alpha$, but we have no quantitative explanation for it.

%\eb{The spectrum is dominated by the accretion disc, no broad absorption features from the white dwarf are visible. Most accreting sources display strong \ion{C}{iv} emission, but here we notice an almost complete} lack of \ion{C}{iv} at 1550\,\AA. \bc{Hmm... I disagree on this: I think we do see a cool WD photosphere, that is, on top, absorbed by the ``iron curtain'' above the disc. Modelling this properly would be a pain (proper WD model with odd abundances plus a curtain model), and I don't know if we can get away with some qualitative statement...}
% \eb{We should try to put a figure on this C-to-N ratio. I will have a look at this and a bit more discussion over the weekend.}
\subsection{\textit{HST} spectrum}
\label{sec:hstspectrum}
\begin{figure}
\centering
\includegraphics[width=\columnwidth]{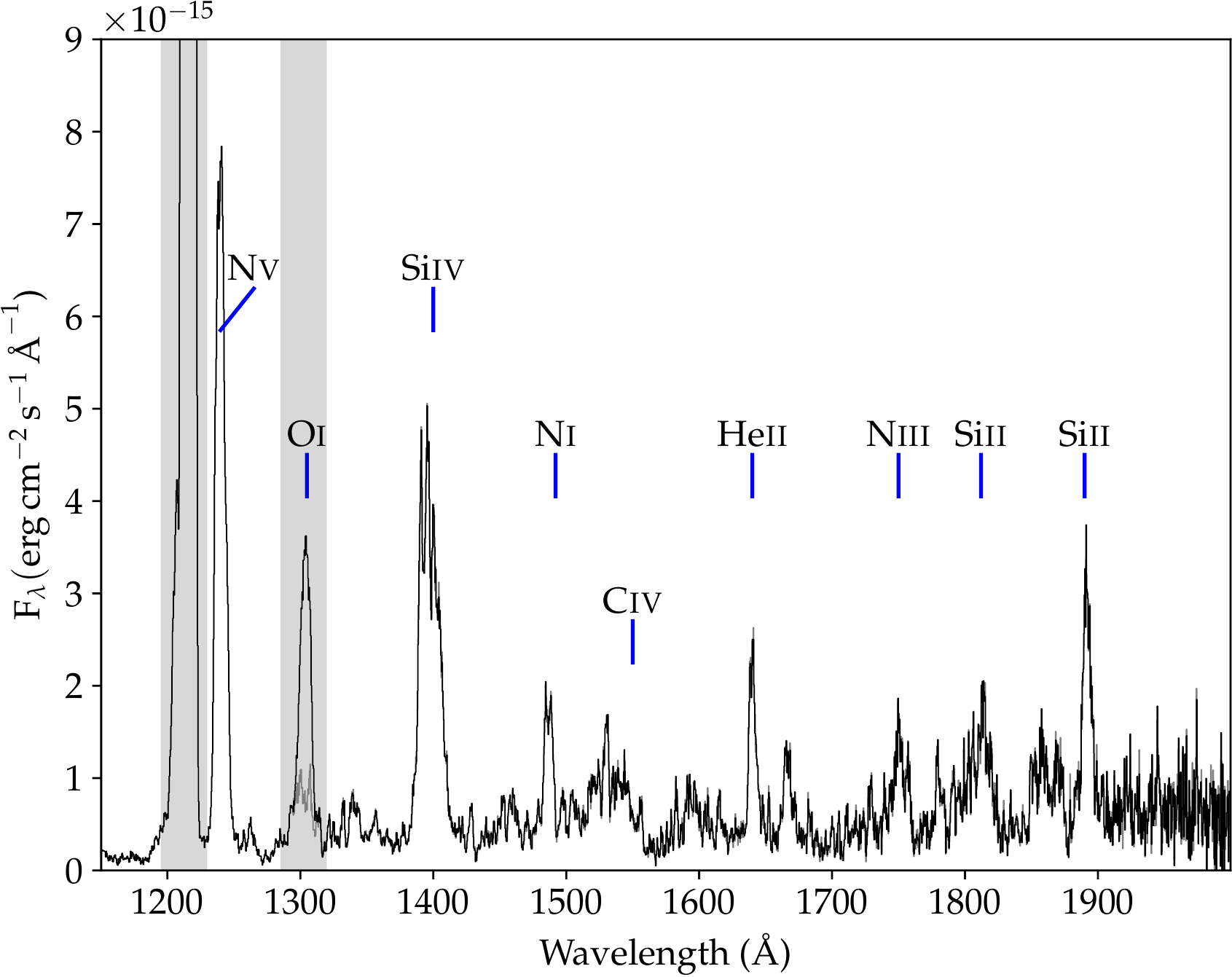}
\caption{The \textit{HST}/COS far-ultraviolet spectrum of \targ.  Most notable is the strong \ion{N}{v} and the complete lack of \ion{C}{iv}. The geocoronal features of Ly$\alpha$ and \ion{O}{i} are marked by grey rectangles. The grey spectrum most notable near the \ion{O}{i} feature at 1295 -- 1314 \AA\ is the average spectrum derived from the exposures taken while \textit{HST} was night-side of the Earth demonstrating that the \ion{O}{i} emission in the main spectrum is largely due to terrestrial airglow.} 
\label{fig:hstspectrum}
\end{figure}

The \textit{HST}+COS spectrum of \targ\ is shown in Fig.~\ref{fig:hstspectrum}.  No clear features from the white dwarf are visible, but the shape of the continuum suggests that it may result from viewing the white dwarf through significant absorption similar to the ``iron curtain'' seen in the dwarf nova OY~Car \citep{Horne1994IronCurtain}. Most accreting sources  display strong \ion{C}{iv}~1550 emission which is not detected in the spectrum of \targ. In addition, there is strong emission from \ion{N}{v} at 1240\,\AA. This suggests an anomalous nitrogen-to-carbon ratio, which was used by \citet{2002MNRAS.337.1105S} in the case of AE~Aqr as an indication that significant CNO processing has taken place in the donor. Therefore the donor star in \targ\ is the descendant of an evolved star, and the white dwarf is accreting material enriched in nitrogen and depleted in carbon due to the CNO cycle. Similar anomalous N/C ratios have been identified in several other CVs, as shown in Table~\ref{tab:evolvedCVs}.  We computed an upper limit on the emission line flux ratio $\ion{N}{v}/\ion{C}{iv}\ga25$, which is even more extreme than those reported for BZ~UMa, EI~Psc, and EY~Cyg by \citet{Gaensicke2003}~--~a consequence of the high signal-to-nose ratio of our COS spectroscopy.

\begin{table}
\caption{Known CVs with anomalous \ion{N}{v} to \ion{C}{iv} emission line ratios and donors that are thought to have experienced some amount of CNO burning.}
\begin{tabular}{l l}
    AE~Aqr    & {\citet{1980MNRAS.191..559J}} \\
    BY~Cam    & {\citet{1987A&A...188...89B}} \\
    TX~Col    & {\citet{1991A&A...250...99M}} \\
    V1309~Ori & {\citet{1996AJ....112..289S}}\\
    MN~Hya    & {\citet{2001ApJ...548..410S}} \\
    EY~Cyg, BZ~UMa & {\citet{2001AAS...199.6201W, 2002AIPC..637...21S}} \\
    GP~Com    & {\citet{1981PASP...93..477L}} \\
              & {\citet{1995MNRAS.274..452M}} \\
    EI\,Psc, CE315   & {\citet{Gaensicke2003}} \\
    U~Sco, V1974~Cyg &  {\citet{2011NewA...16...19S}} \\
    CSS 120422:J111127+571239 & {\citet{2015ApJ...815..131K}} \\
    \targ\     & This paper \\
\end{tabular}
\label{tab:evolvedCVs}
\end{table}

\subsection{Spin of the white dwarf}
\label{sec:hstphotometry}
\begin{figure}
\centering
\includegraphics[width=\columnwidth]{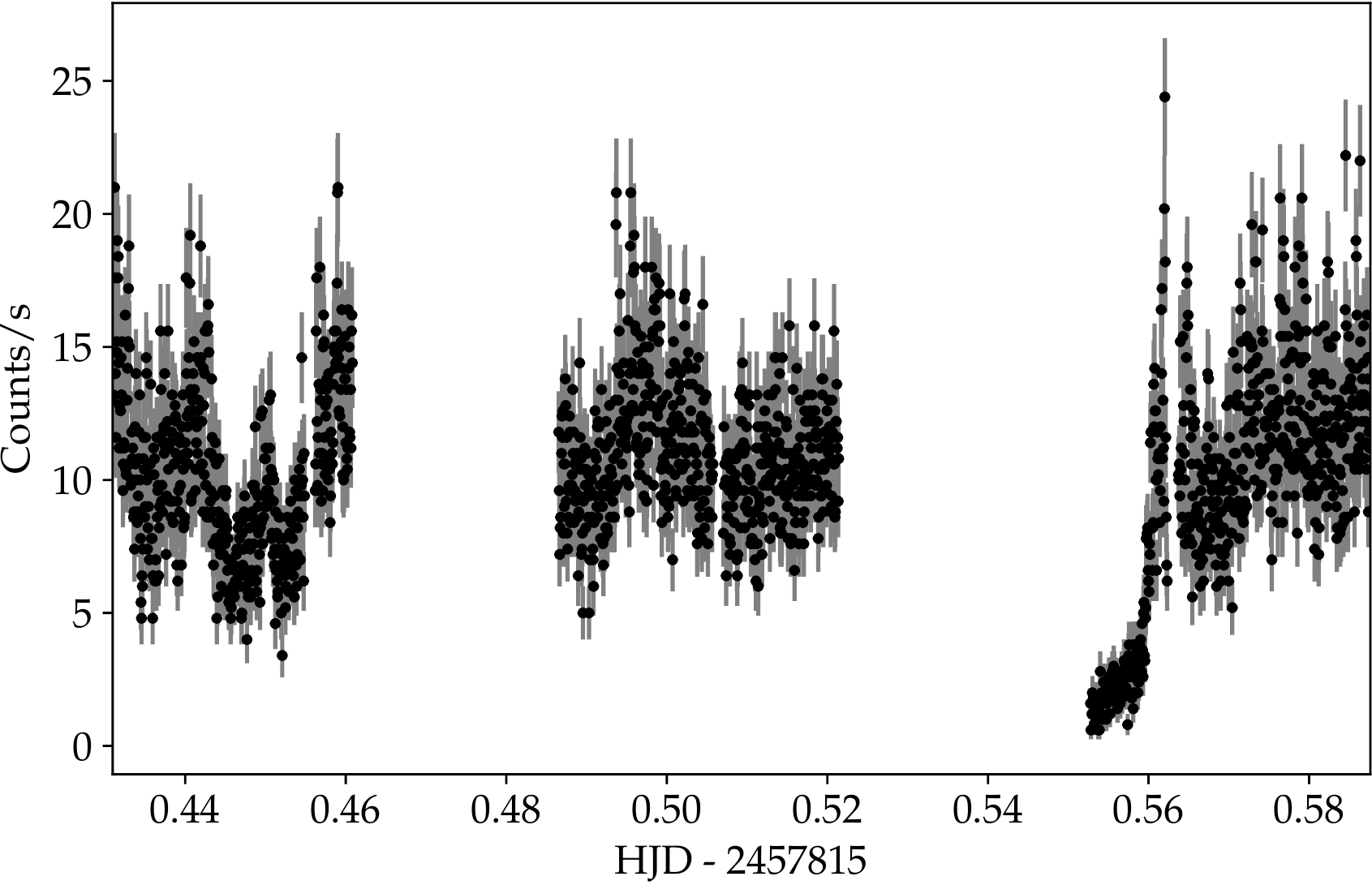}
\caption{Photometry derived from the \textit{HST} spectroscopy as described in the text. The large scatter in the data is partially due to a true variability in the UV flux over a 38.9\,s period.}
\label{fig:hst_lightcurve}
\end{figure}

As discussed in section \ref{HSTobservations}, the \textit{HST} ultraviolet spectroscopy was also used to derive time-resolved ultraviolet photometry. Our data covered approximately 75 per\,cent of an orbit and included a small section and the egress of a primary eclipse. The photometry is shown in Fig.~\ref{fig:hst_lightcurve}. The amount of scatter seen in the data in phases away from eclipse is significantly larger than expected given the uncertainties from photon statistics. 

\begin{figure}
\centering
\includegraphics[width=\columnwidth]{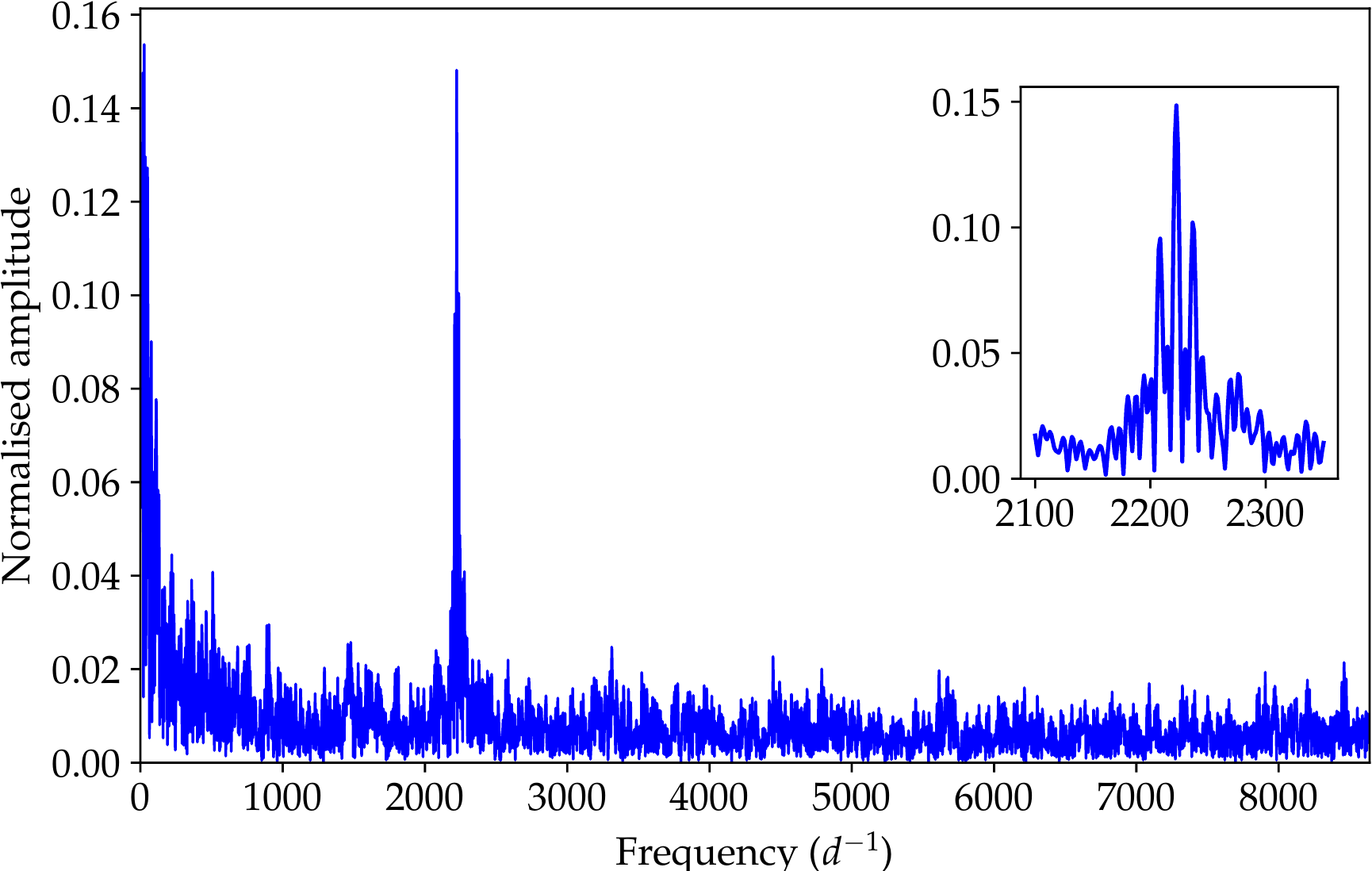}
\caption{The fractional amplitude spectrum up to the Nyquist frequency for our adopted $5$\,s
sampling rate. Data taken during the eclipse do not show the 38.9\,s modulation and have been excluded from this analysis.}
\label{fig:hst_LS}
\end{figure}

\begin{figure}
\centering
\includegraphics[width=\columnwidth]{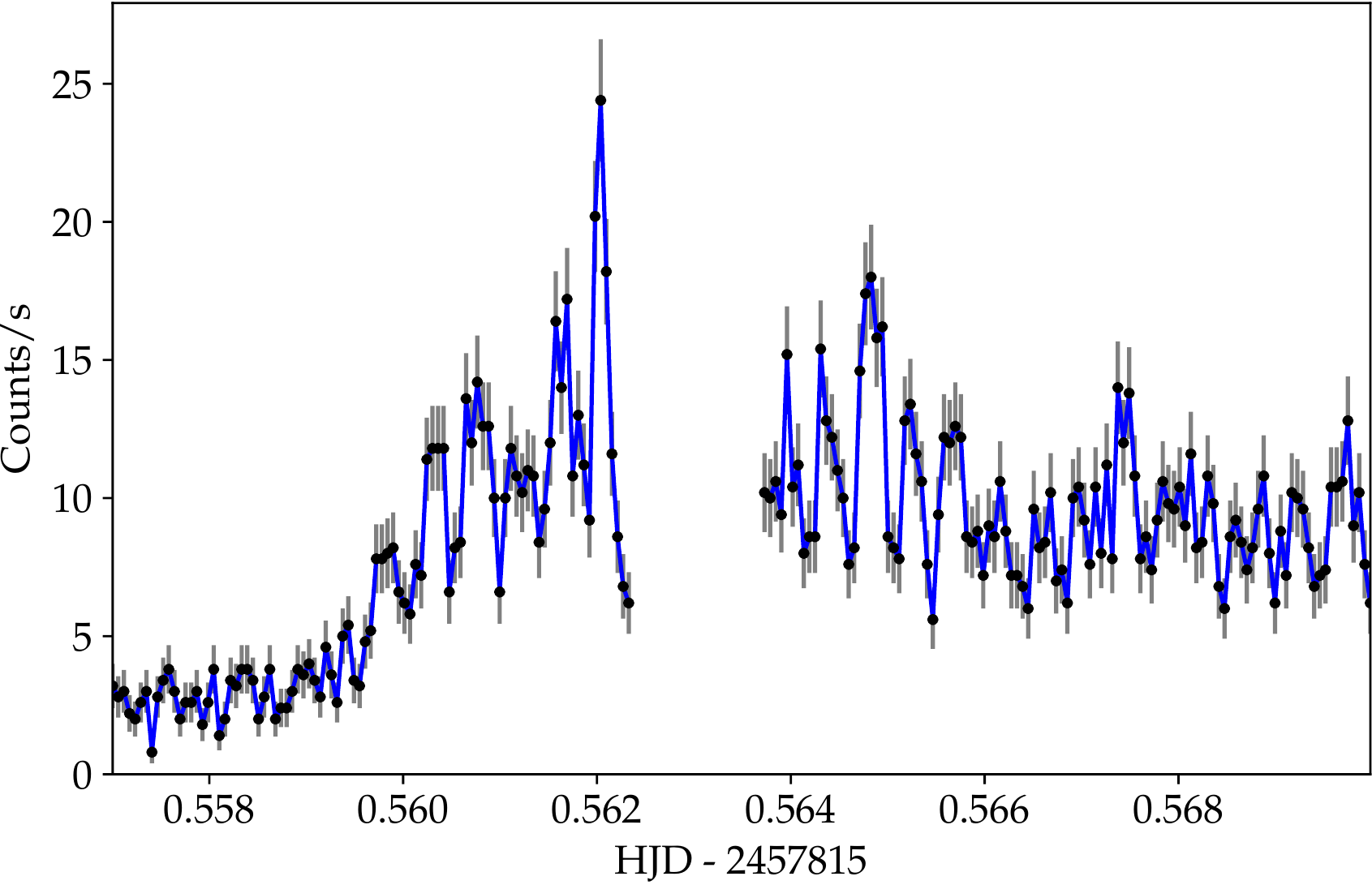}
\caption{Closer view of some of the \textit{HST} derived UV photometry. This shows the eclipse egress and the 38.9\,s modulation which turns on during the egress.}
\label{fig:hst_zoom}
\end{figure}

In order to look for pulsations from the white dwarf, we computed a periodogram of the \textit{HST} data. The fractional amplitude spectrum is shown in Fig.~\ref{fig:hst_LS}. There is a strong peak  at around 2222 cycles\,d$^{-1}$ caused by a periodic modulation with a semi-amplitude of 15 per\,cent and a period of  $38.875 \pm 0.005$\,s. This period was derived by fitting a sinusoid to the modulation with the peak of the periodogram as a starting point. The uncertainty in the period is derived from the covariance matrix of the least squares fit. A closer inspection (Fig.~\ref{fig:hst_zoom}) of the light curve shows that the modulation is directly visible, and that it disappears during the eclipse. The source of this modulation, which was observed during quiescence, can only be due to the spin of the underlying white dwarf, or possibly its orbital sideband. The spin period of the white dwarf is then either equal to the observed $38.875$\,s period, or it is very close to it. It is possible that we are seeing an identical pattern generated by each pole in which case the true spin period is in fact twice 38.875\,s.  There are no other significant periodic signals, but note that the \textit{HST} data covers less than an orbit of \targ\ and therefore cannot resolve signals spaced by the orbital frequency or less from each other. The ephemeris of this modulation is
\begin{equation}\mathrm{BJD(TDB)} = 2457815.43095(8) + E \times 4.4994(4)\times 10^{-4},\end{equation}
where the constant term marks the time of maximum flux.

\begin{figure}
\centering
\includegraphics[width=\columnwidth]{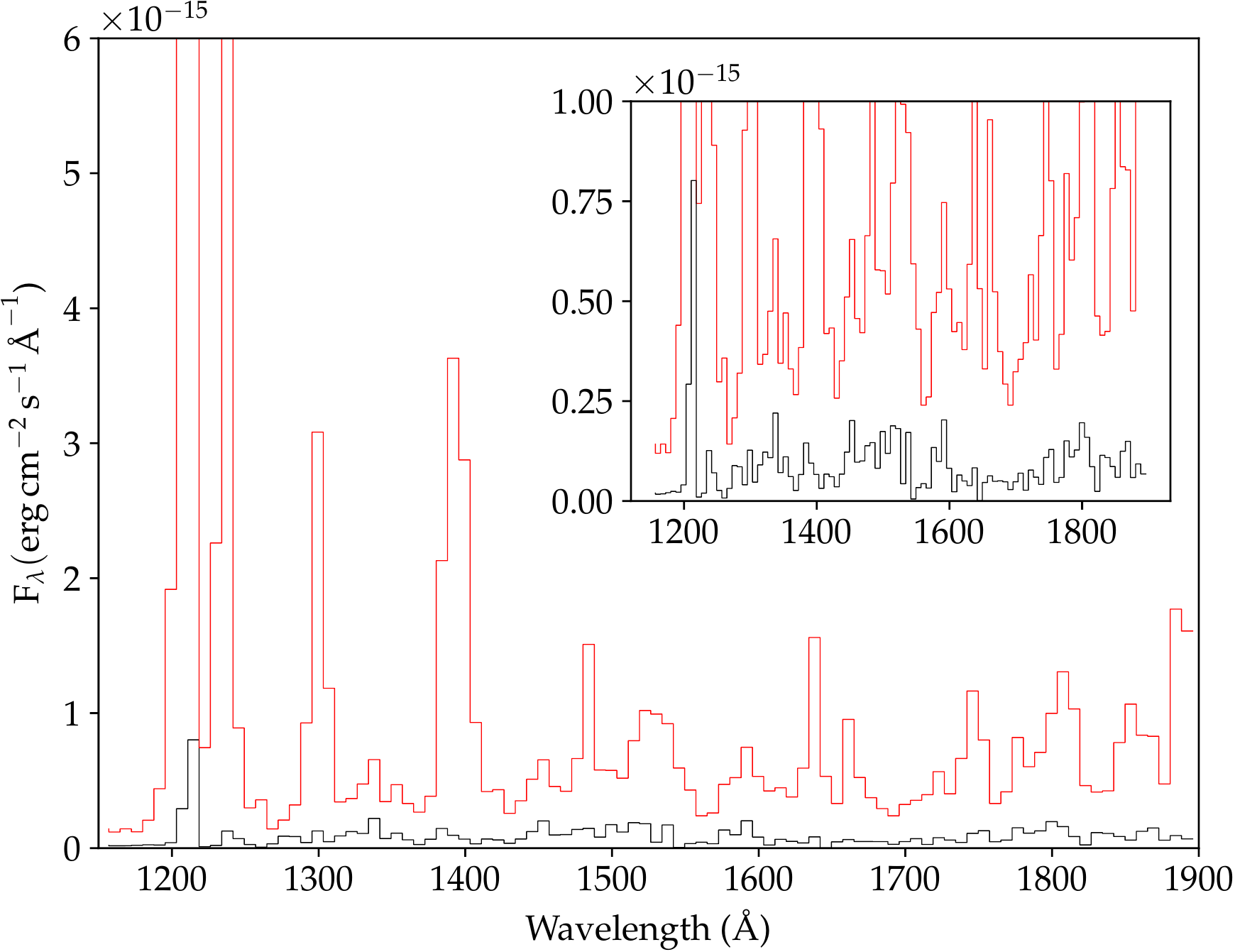}
\caption{The pulsation spectrum (shown in red) along with the overall average flux (larger values in black). The pulsation amplitude comes from the continuum region of the spectrum and not from the emission features, as can be observed near the \ion{Si}{iv} feature at 1400\,\AA. }
\label{fig:pulsespectrum}
\end{figure}

By time-binning individual data from the CORRTAG files written by the COS instrument, we were able to generate time resolved spectra to sample the pulse signal of this 38.9\,s modulation. We generated 2132 spectra with 3.9\,s time resolution and then averaged them into 10 phase bins. Then, after rebinning the wavelength axis into 100 wavelength bins, we computed the amplitude of the pulsation as a function of wavelength. The pulsation spectrum is shown in Fig.~\ref{fig:pulsespectrum}.  It is clear that the pulsations emanate from the flux in the continuum rather than from the emission lines, and the pulsation spectrum appears to be somewhat bluer (hotter) than the mean spectrum. In the study of the rapidly rotating white dwarf primary in AE~Aqr, \citet{Eracleous1994} showed that their pulsation spectrum was matched by a model of a solar abundance white dwarf with $\log g = 8$ and conclude that the pulsations are the consequence of the heated pole caps on the surface of the white dwarf.  We do not have sufficient data to perform a similar analysis, but we assume that the modulation in \targ\ is due to a similar process.  

\section{Discussion}
The white dwarf mass, $M_1 = 0.87\,\msun$, is consistent with the average mass of CV white dwarfs \citep{2011A&A...536A..42Z}. In contrast, the donor star's mass of $0.29\,\msun\,$ (Table~\ref{tab:ephem}) is far less than the expected mass of a K5 star on the main sequence, $0.68\,\msun$ \citep{2013ApJS..208....9P} and is closer instead to the typical mass of an M3.5 donor star, \citep{2006MNRAS.373..484K}. Moreover, the radius of a $0.29\,\msun$ main-sequence star is expected to be around $0.30\,\rsun$ \citep{2013ApJS..208....9P}, so the donor here is $\approx 50$~per cent larger than expected. It is clear that the donor is overly large and hot compared to main-sequence stars of similar mass, a consequence presumably of past mass loss that has happened on a timescale shorter than its Kelvin-Helmholtz (thermal) timescale of $\tau_\mathrm{KH} \sim 7 \times 10^7\,$yr. The \textit{HST} spectra reveal anomalous abundance ratios between \ion{N}{V}, \ion{Si}{IV} and \ion{C}{IV}; in fact we find no evidence for C in our spectra. This leads us to conclude that \targ\ is accreting from an evolved donor that has undergone CNO processing. One possible explanation for the short time scale of mass transfer and also the evidence for CNO burning is that \targ\ may have started accretion from a donor that was more massive than the white dwarf leading to a high rate of mass transfer and significant stellar evolution \citep{2011NewA...16...19S, 2002MNRAS.337.1105S}. \citet{2015MNRAS.452.1754P} identified TYC~6760-497-1 as the first potential progenitor of a system that will undergo thermal time scale transfer during its evolution, and will end up with an inverted NV/CIV emission line flux ratio. \citet{Podsiadlowsk2003} use binary population synthesis (BPS) techniques to show that, according to their models, CVs with a periods >5\,hrs are likely to be dominated by such evolved donors. 

%Schenker & King (2002) show the evolution of the donor mean density as a function of its mass (their Fig. 2) for boththe ZAMS and evolved cases. With an orbital period of 5.0\,h and a density of $4.5 g cm^{−3}$, \targ\ appears to lie along the track of the case of strong thermal-timescale mass transfer, meaning that 

Many dwarf novae with periods similar to \targ, undergo outbursts several times per year, eg RX~And $\sim 15\,$d, EX~Dra $\sim 25\,$d CZ~Ori $\sim 40$\,d and AT~Cnc $\sim 30$\,d. It is thought that this outburst frequency is driven by the donor stars experiencing strong magnetic stellar wind braking driving the evolution towards shorter orbital periods, resulting in moderately high mass flow rates through the discs in these systems. Looking at the archival photometric data (section \ref{sec:archive}) however, \targ\ only seems to outburst about once every ten years, assuming that we have not missed any outbursts when the target was seasonally inaccessible. This suggests that the mass transfer rate is lower than predicted by standard CV evolution models. This may be a consequence of the unusual state of the donor star which should be shrinking towards a more main-sequence-like structure, driving it towards detachment from its Roche lobe.  Alternatively, or perhaps additionally,  the low outburst frequency may be the result of disruption of the inner disc by the magnetic white dwarf, as suggested by \citet{1989MNRAS.238..697A}. \citet{2017A&A...602A.102H} show that dwarf nova outbursts from intermediate polars are expected to be less frequent and of shorter duration than those from non-magnetic CVs. It appears then that \targ\ could be a marginal candidate for this group, with outbursts every 10 years, lasting about 15 days. However, two other CVs with evolved donors, HS0218+3229 \citep[$P_\mathrm{orb} = 7$\,h]{2009A&A...496..805R} and KIC5608384 \citep[$P_\mathrm{orb} = 9$\,h]{2019MNRAS.489.1023Y} also outburst infrequently and are not obviously magnetic, perhaps suggesting that it is the state of the donor star that is of most importance.

The unusual state of the donor star points to relatively recent emergence from a high rate of mass transfer. \citet{2002MNRAS.337.1105S} developed a model in which AE~Aqr emerged from a supersoft X-ray phase during which the accretion rate was high enough ($\sim 10^{-7}\,\msun\,\text{yr}^{-1}$) to drive steady fusion on the surface of the white dwarf. One would also expect the white dwarf to spin up during this phase. Once the high rate of mass transfer ceases, there would be tendency for the white dwarf to spin down, and \citet{2002MNRAS.337.1105S} suggest that this explains why we now see AE~Aqr in a state of rapid spin-down of the white dwarf, which is losing more rotational energy than seen from the system as a whole. \targ\ seems in every way to be a second example of this process. Not only does it have a clearly evolved donor star, but it contains a rapidly spinning white dwarf, which is only a little slower than the white dwarf in AE~Aqr itself. It will be fascinating to see whether, like AE~Aqr and the possibly-related system AR~Sco \citep{MarshARSco, 2018AJ....156..150S}, the white dwarf in \targ\ is in a state of rapid spin-down. This will require alias-free measurement of the pulsation phase over a number of years, not possible from the data we have in hand at present. \targ\ differs from AE~Aqr in having a normal-looking accretion disc, which is visible in quiescence through line emission, and capable of driving outbursts, albeit rarely. AE~Aqr on the other hand is known as the only propeller-type system amongst accreting white dwarfs, flinging matter transferred from the donor out of the system altogether as proposed by \citet{1997MNRAS.286..436W}. AR~Sco is more extreme still and seems to have entirely ceased to accrete \citep{MarshARSco}, possibly as a result of the injection of spin angular momentum into the orbit. In this picture, \targ\ is a good candidate for the less extreme end, with a magnetic field too weak to disrupt its accretion disc let alone lead to propeller ejection and the complete cessation of accretion. The recently reported CTCV~J2056-3014 \citep{2020ApJ...898L..40L} is an intermediate polar (IP) with a white dwarf spin period of 29.6\,s and this system looks to be currently the fastest confirmed spin in a CV.  CTCV~J2056-3014 is thought to have a lower magnetic field strength than is typical for IPs and a much shorter period (1.76\,h) than usual. A study of this system's donor would be particularly interesting as it could form part of this new family of CVs with evolved donors and rapid spin of the white dwarf.

It is interesting that with \targ, another system in addition to AE~Aqr has been found with a rapidly-spinning white dwarf that seems to have resulted from a recent phase of high accretion. If there is a puzzle associated with white dwarf spin evolution under accretion, it is perhaps why more of them are not found to be rapidly spinning, because only a little mass needs to be accreted to bring a white dwarf close to its break-up spin, much faster than many accreting white dwarfs are found to rotate. \citet{1998ApJ...505..339L} suggested that angular momentum loss during classical nova eruptions might be the reason for slow spin rates. This fits remarkably well with the recent supersoft phase idea, because a key feature of the supersoft phase is that the accreted hydrogen steadily fuses on the white dwarf surface so that there are no classical nova eruptions in these systems. This might suggest that rapidly spinning white dwarfs are more likely to be located in systems with evolved donor stars, a testable hypothesis.

\section{Conclusions}

We find that \targ\ belongs to a select group of CVs in which the donor star is significantly over-luminous and over-sized (by 50~per cent) for its mass. This indicates that it is probably the remnant of a phase of high rate mass transfer. It is moreover eclipsing and shows phenomena in its spectra associated with both the eclipse of its disc by the donor star and the eclipse of the donor star by the disc. Most remarkably of all, \targ\ shows strong pulsations on a period of 39\,s in \textit{HST} ultraviolet data. This is a clear sign of a rapidly spinning white dwarf, and amongst the fastest known amongst CVs such as AE~Aqr and the recently reported CTCV~J2056-3014. AE~Aqr also hosts an evolved donor star, and these two systems may share a history of white dwarf spin-up through high rate accretion. We speculate that this may only occur if accretion at a rate high enough to suppress classical nova eruptions has taken place.

\section*{Acknowledgements}
TRM, BTG \& OT acknowledge support from the Science and Technology Facilities Council (STFC) grant numbers ST/P000495/1 and ST/T000406/1 and EB from STFC grant number ST/S000623/1. This paper makes use of data from the first public release of the WASP data \citep{2010A&A...520L..10B} as provided by the WASP consortium and services at the NASA Exoplanet Archive, which is operated by the California Institute of Technology, under contract with the National Aeronautics and Space Administration under the Exoplanet Exploration Program. The research leading to these results has received funding from the European Research Council under the European Union's Seventh Framework Programme (FP/2007-2013) / ERC Grant Agreement n.320964 (WDTracer). We acknowledge the ongoing support of the Instituto de Astrof\'{i}sica de Canarias that has enabled our work. This research has made use of the SIMBAD database, operated at CDS, Strasbourg, France.  

The authors would like to thank the astronomers of the AAVSO network who provided excellent photometry of \targ\ during the outburst and in particular Paul Benni, Ivaldo Cervini and Thomas Wikander whose support during the \textit{HST} observations helped to ensure the safety of the spacecraft. We would also like to thank Ian Skillen and Ovidiu Vaduvescu at the Isaac Newton Group in La Palma for their invaluable WHT ISIS spectra taken during four nights of service time. Jorge Casares very kindly sent us his list of template spectra of K-type stars that helped in identifying the spectral type of the secondary and allowed us to calculate the rotational broadening. 

We would also like to thank the anonymous referee for their helpful, insightful and constructive feedback. 

\section*{Data availability}
The data underlying this article will be shared on reasonable request to the corresponding author.

%%%%%%%%%%%%%%%%%%%%%%%%%%%%%%%%%%%%%%%%%%%%%%%%%%



\bibliographystyle{mnras}
\bibliography{references} % if your bibtex file is called example.bib

%%%%%%%%%%%%%%%%%%%%%%%%%%%%%%%%%%%%%%%%%%%%%%%%%%

%%%%%%%%%%%%%%%%% APPENDICES %%%%%%%%%%%%%%%%%%%%%

%\appendix
%\section{Additional figures}

% Don't change these lines
\bsp	% typesetting comment
\label{lastpage}
\end{document}